\documentclass[a4paper,12pt]{article}
\usepackage{geometry}
\usepackage{setspace}

\usepackage{enumitem}
\usepackage{amsmath,amssymb,amsthm}
\usepackage{mathtools}

\newtheorem{theorem}{Theorem}
\newtheorem{lemma}{Lemma}
\newtheorem{proposition}{Proposition}
\newtheorem{corollary}{Corollary}
\theoremstyle{definition}
\newtheorem{assumption}{Assumption}

\numberwithin{subassumption}{assumption}

\allowdisplaybreaks

\renewcommand{\d}{\mathrm{d}}
\newcommand\T{{ \mathrm{\scriptscriptstyle T} }}
\renewcommand{\Pr}{\mathrm{pr}}
\newcommand{\indep}{\mathrel{\reflectbox{\rotatebox[origin=c]{90}{$\models$}}}}
\makeatletter
\newcommand*\dott{\mathpalette\dott@{.75}}
\newcommand*\dott@[2]{\mathbin{\vcenter{\hbox{\scalebox{#2}{$\m@th#1\bullet$}}}}}
\makeatother

\usepackage{booktabs}
\usepackage{multirow}
\usepackage{array}

\usepackage{tikz}
\usetikzlibrary{arrows.meta,automata,calc,positioning,shapes.arrows,shapes.geometric,shapes.multipart,decorations.pathreplacing,decorations.pathmorphing,positioning,swigs}
\tikzset{vertex/.style={inner sep=0pt,minimum size=1em},
         sq/.style={draw,rectangle,inner sep=0pt,minimum size=1em},
         hidden/.style={draw,shape=circle,preaction={fill=gray!50,fill opacity=0.5},inner sep=1pt,minimum size=1em},
         swig vsplit={gap=3pt,line color right=red},
         ell/.style={draw,inner sep=0pt,shape=ellipse}}
\tikzset{snake it/.style={decorate, decoration={snake, amplitude=.4mm,segment length=1.5mm,post length=1mm,pre length=1mm}}}

\usepackage{natbib}

\usepackage{multibib}
\newcites{suppmat}{References}
\bibliographystylesuppmat{apalike}

\usepackage{times}
\usepackage{newtxmath}
\usepackage[cal=cm,bb=ams]{mathalpha}

\usepackage{authblk}

\usepackage{titlesec}
\titleformat*{\section}{\bfseries}
\titleformat*{\subsection}{\bfseries}
\titleformat*{\paragraph}{\bfseries}

\usepackage{minitoc}
\noptcrule

\newcommand\E{E}
\renewcommand\u{u}
\renewcommand\H{\mathcal{H}}
\newcommand\Q{\mathcal{Q}}
\newcommand\s{g}
\renewcommand\c{c}
\renewcommand\P{\mathcal{P}}

\usepackage{siunitx}
\usepackage{etoolbox}

\makeatletter
\patchcmd{\@maketitle}{\large}{\normalsize}{}{}
\patchcmd{\@maketitle}{\LARGE}{\large}{}{}
\makeatother

\title{Proximal indirect comparison}
\date{}

\author[1]{Zehao Su\footnote{Corresponding author.}}
\author[1]{Helene Charlotte Rytgaard}
\author[2]{Henrik Ravn}
\author[1]{Frank Eriksson}
\affil[1]{Section of Biostatistics, University of Copenhagen, Copenhagen, Denmark}
\affil[2]{Novo Nordisk A/S, S\o{}borg, Denmark}

\begin{document}

\maketitle

\begin{abstract}
  We consider the problem of indirect comparison, where a treatment arm of interest is absent by design in one randomized controlled trial but available in the other.
  The former is the target trial, and the latter is the source trial.
  The identifiability of the target population average treatment effect often relies on conditional transportability assumptions.
  However, it is a common concern whether all relevant effect modifiers are measured and controlled for.
  We give a new proximal identification result in the presence of shifted, unobserved effect modifiers based on proxies: an adjustment proxy in both trials and an additional reweighting proxy in the source trial.
  We propose an estimator which is doubly-robust against misspecifications of the so-called bridge functions and asymptotically normal under mild consistency of estimators for the bridge functions.
  We use two weight management trials as a context to illustrate selection of proxies and apply our method to compare the weight loss effect of active treatments from these trials.
  
\end{abstract}

\section{Introduction}

Indirect comparison is the contrast of treatments that are not compared in head-to-head randomized controlled trials (RCTs).
An important application of indirect comparison in health technology assessment is the comparison of a new treatment and an existing treatment in the health system, when both treatments are only studied in placebo-controlled RCTs.
Indirect comparison between the two active treatments can be made through the shared placebo arm, if the effect of the existing treatment against placebo can be evaluated in the study population of the RCT investigating the new treatment.
In this sense, indirect comparison can be viewed as an instance of transportability in causal inference.
Transportability of the effect of the existing treatment versus placebo (or the lack of such transportability) from the source to the target RCT determines whether the effect between the new and the existing treatment can be established.
It is therefore hardly surprising that current methods for indirect comparison require effect-measure transportability, adjusting for shifted effect modifiers \citep{colnet2024causal}, that is, the effect modifiers that do not follow the same distribution across the RCTs.
However, when there are unobserved shifted effect modifiers, transportability cannot be established by controlling for the observed baseline variables.
The lack of transportability jeopardizes the external validity of treatment effects in an indirect comparison.
If the treatments of interest come from RCTs which are conducted with a considerable time gap apart, there may be changes in the standard of care that could affect the treatment effects.
Social determinants of health, which are often unmeasured in RCTs, can also change the magnitude of treatment effects.

In this work, we use negative controls, or proxies, to minimize external validity bias.
In observational studies, negative controls are known to help detect unmeasured confounding \citep{lipsitch2010negative}.
Recently, a family of methods called proximal causal inference has shown how appropriately selected proxies may rectify confounding bias.
\citet{miao2018identifying} demonstrated a nonparametric identification formula for the counterfactual distribution of outcomes with a pair of complementary proxies in the presence of unmeasured confounders.
Under a nearly identical design, \citet{cui2024semiparametric} proposed a proximal doubly robust estimator for average treatment effect based on two identification strategies.

In indirect comparisons, randomization eliminates the threat of confounding, and we use proxies to tackle bias arising from shifted, unobserved effect modifiers.
We propose a novel method that extends proximal causal inference to indirect comparison when individual patient data is available in all RCTs.
Our proposed estimator also relies on a pair of proxies, namely a reweighting proxy and an adjustment proxy.
The proxies correspond to identification strategies that mirror participation odds weighting and the g-formula for transporting causal effects.
We remark that while both proxies are required in the source RCT, only the adjustment proxy needs to be collected in the target RCT.
Our proposed estimator handles both continuous and binary outcomes.
We show that the estimator is robust against misspecifications of the required nuisance functions of the proxies.

\section{Indirect comparison with unmeasured shifted effect modifiers}
\label{sec:problem}

\begin{figure}
  \centering
  \includegraphics[scale=1]{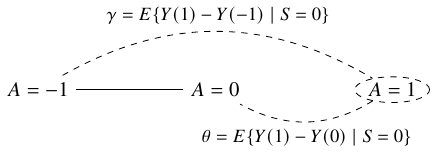}
  
  \caption{Indirect comparison parameters defined on the target RCT population \(S=0\).
    Treatment \(A=1\) is unavailable in the target RCT.
  }
  \label{fig:comparison}
\end{figure}

Consider two treatment pairs, \(A\in\{0,1\}\) and \(A\in\{0,-1\}\), where the former is the treatments investigated in the source RCT \(S=1\) and the latter is those in the target RCT \(S=0\).
This corresponds to the situation where there is a treatment shared by the two RCTs, in this case \(A=0\), so that the comparison of the treatments \(A=1\) and \(A=-1\) may be made with the help of the common treatment arm.
Indirect comparison of this kind is called anchored comparison \citep{phillippo2018methods}.
The setup described here is a subset of network meta-analysis, which usually places interest on the causal effects comparing at least three different pairs of interventions.
The treatment \(A=1\) that we wish to emulate in \(S=1\), also referred to as the missing treatment, can be a placebo or an active treatment.
Additionally in each RCT a set of baseline covariates \(X\) is measured.
Let \(Y(a)\) denote the potential outcome under the intervention \(A=a\in\{-1,0,1\}\).
In this paper we consider as the target parameter \(\theta\) the average treatment effect (ATE) in the target population comparing treatments \(A=1\) and \(A=0\); that is, \(\theta=\E\{Y(1)-Y(0)\mid S=0\}\).
The indirect comparison parameter \(\gamma=\E\{Y(1)-Y(-1)\mid S=0\}\) is the difference between \(\theta\) and the ATE in the target RCT, \(\E\{Y(-1)-Y(0)\mid S=0\}\).
Figure~\ref{fig:comparison} contains a diagram describing the parameters \(\gamma\) and \(\theta\).
For the identifiability of \(\theta\), the natural effect measure for transportability is the conditional average treatment effect (CATE).
The mean scale is a common choice in various problems where the outcome is continuous (e.g., body weight) or binary (e.g., occurrence of a cardiac arrest).

The fundamental problem of indirect comparison is that the treatment \(A=1\) is never observed in the target RCT \(S=0\).
In order to establish identifiability for the target parameter in the observed data, we need a set of plausible assumptions which justify the transportability of the treatment-specific mean from the source population to the target population.
The existing indirect comparison and network meta-analysis literature tend to assume effect-measure transportability \citep{phillippo2018methods,phillippo2020multilevel}, which does not hold when conditioning on the observed baseline covariates cannot completely account for differences in the effect between RCTs, because there exist unmeasured shifted effect modifiers.
In hope of capturing all effect modifiers, it may be tempting to adjust for as many baseline or pre-treatment variables as possible in order to achieve valid effect-measure transportability.
However, this strategy is not foolproof, as it may instead result in additional M-bias \citep{cinelli2024crash}.

We propose a weaker version of CATE transportability conditioning on both the baseline covariates \(X\) and the unobserved effect modifiers \(U\).

\begin{assumption}
  \label{asn:causal}
  \(\Pr(X,U\mid S=0)\)-almost surely:
  \begin{enumerate}[label=(\roman*)]
  \item\label{asn:consistency}(Consistency) \(Y(a)=Y\) whenever \(A=a\in\{0,1\}\);
  \item\label{asn:randomization}(Randomization) \(\{Y(1),Y(0),U\}\indep A\mid (X,S=1)\);
  \item\label{asn:transportability} (CATE transportability) \(\E\{Y(1)-Y(0)\mid X,U,S=0\}=\E\{Y(1)-Y(0)\mid X,U,S=1\}\);
  \item\label{asn:positivity} (Positivity) \(\Pr(S=1\mid X,U)\Pr(A=a\mid X,S=1) > 0\) for \(a\in\{0,1\}\).
  \end{enumerate}
\end{assumption}

Assumption~\ref{asn:causal}\ref{asn:consistency} states that the observed outcome is the potential outcome under the assigned treatment.
Assumption~\ref{asn:causal}\ref{asn:randomization} is a consequence of treatment randomization that possibly conditions on the baseline covariates.
Assumption~\ref{asn:causal}\ref{asn:transportability} requires that the treatment effect of \(A=1\) versus \(A=0\) is transportable after conditioning on both \(X\) and \(U\).
Assumption~\ref{asn:causal}\ref{asn:positivity} is crucial for the transportability of causal effects from the source population \(S=1\) to the target population \(S=0\).
It requires that all values of \((X,U)\) existing in the target population must also be observable in the source population.
The positivity assumption on trial participation essentially guarantees no extrapolation of information from the source trial.
As we will see in \S\ref{sec:proxy}, violation of positivity not only compromises the identifiability of the target parameter but also has great implications on its estimation with observed data.

The causal assumptions give rise to identifiability of the target parameter in the full data distribution.
Let \(\alpha=\Pr(S=0)\) be the probability of belonging to the target RCT for any individual.
We denote the propensity score in the source trial by \(e(a\mid X)= \Pr(A=a\mid X,S=1)\) for \(a\in\{0,1\}\).
In the source RCT, we also let \(\tilde{Y}=(2A-1)Y/e(A\mid X)\) denote a transformed outcome, whose behavior is comparable to that of the contrast \(Y(1)-Y(0)\), in the sense that their conditional expectations on \((X,U)\) are equal.

\begin{proposition}[Latent identifiability]
  \label{ppn:identification-unmeasured}
  Suppose Assumption~\ref{asn:causal} holds.
  The target parameter is identifiable as
  \begin{equation}
    \theta = \E\bigl\{\E(\tilde{Y}\mid X,U,S=1)\,\big\vert\, S=0\bigr\}.\label{eqn:adjustment}
  \end{equation}
  Equivalently,
  \begin{equation}
    \theta = \frac{1}{\alpha}\E\bigg\{S\frac{\Pr(S=0\mid X,U)}{\Pr(S=1\mid X,U)}\tilde{Y}\bigg\}.\label{eqn:reweighting}
  \end{equation}
\end{proposition}

The proof of Proposition~\ref{ppn:identification-unmeasured} can be found in Supplementary Material \S{}\ref{sec:identification-unmeasured-proof}.
These identification formulae of \(\theta\) are very similar to the transportability results in Theorem 1 of \citet{dahabreh2023efficient}, where the outcome model and the participation odds depend on the baseline covariates \(X\) only.
The distinction mainly lies in their assumptions for the exchangeability of trial participation and subsequently the exchangeability of treatment assignment, which they assume to hold without the additional unobserved covariates \(U\).

Proposition~\ref{ppn:identification-unmeasured} suggests that in order to identify the target parameter \(\theta\), we need at least the knowledge of either the mean outcome difference \(\E(\tilde{Y}\mid X=x,U=u,S=1)\) or the trial participation odds \(\Pr(S=0\mid X=x,U=u)/\Pr(S=1\mid X=x,U=u)\).
However, both quantities depend on the unobserved effect modifiers \(U\), and neither would be identifiable without further assumptions and/or extra information.

\section{Reweighting and adjustment proxies}
\label{sec:proxy}

\subsection{Proximal identifiability}

In this section, we extend the proximal causal inference framework to handle scenarios in indirect comparison where CATE transportability fails to hold conditionally on the baseline covariates \(X\).
We refer to this approach as proximal indirect comparison.
The core idea of proximal indirect comparison is that the knowledge of a pair of proxies \((Z,W)\) helps to learn the underlying dependence of \(Y(1)-Y(0)\) on \(U\), thereby restoring the identifiability of the target parameter using observed data.
We call \(Z\) the reweighting proxy and \(W\) the adjustment proxy.
The intuition is that \(Z\) will be used to reweight the samples from the source RCT to match the composition of the target RCT like in \eqref{eqn:reweighting}, and \(W\) will appear in the adjustment formula for transportability in \eqref{eqn:adjustment}, as if we had adjusted for \(U\).

Before discussing identifiability of the parameter, we formally introduce the variables at our disposition.
Specifically, we observe \(O^1=(A,X,Y,W,Z)\) in the source RCT and \(O^0=(X,W)\) in the target RCT, so the observed variables can be represented by \(O=(S,SA,X,SY,W,SZ)\).
We describe the following distinction between the observed data and the full data.
Denote the full data distribution over \((O,U)\) by \(P^{\u}\).
We can marginalize it to a measure \(P\) over \(O\).
Further, \(P^{1}\) stands for the conditional probability measure over \(O^{1}\) such that \(P^{1}(O^{1}\in \cdot)=P(O^{1}\in \cdot, S=1)/P(S=1)\).

We require that the proxies satisfy a set of conditional independences.

\begin{assumption}[Adjustment and reweighting proxies]
  \label{asn:identify}
  In \(P^{\u}\):
  \begin{enumerate}[label=(\roman*)]
  \item\label{eqn:identify-a} \((Z,W,U)\indep A\mid (X,S=1)\);
  \item\label{eqn:identify-zw} \(Z \indep W\mid (X,U,S=1)\);
  \item\label{eqn:identify-zy} \(Z \indep Y\mid (A,X,U,S=1)\);
  \item\label{eqn:identify-ws} \(W \indep S\mid (X,U)\).
  \end{enumerate}
\end{assumption}

\begin{figure}
  \centering
  \includegraphics[scale=1]{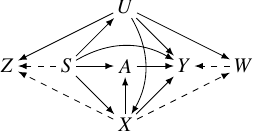}
  \caption{A DAG compatible with Assumption~\ref{asn:identify}.
    Dashed edges can be removed.}
  \label{fig:hidden}
\end{figure}

This assumption describes the relations between valid proxies and other variables.
In the source RCT, Assumptions~\ref{asn:identify}\ref{eqn:identify-a}--\ref{asn:identify}\ref{eqn:identify-zy} essentially require that randomization does not depend on the proxies, that the proxies are not associated through other unmeasured variables besides \(U\), and that \(Z\) has no causal effect on \(Y\), unless it is mediated through \((X,U)\).
Also, if \(Z\) and \(W\) occur after randomization, they must be negative control outcomes in the sense that they can share many causes with \(Y\) but may not be caused by the treatment \(A\).
Across the two RCTs, Assumption~\ref{asn:identify}\ref{eqn:identify-ws} requires that the difference in the distribution of \(W\) is totally accounted for by \((X,U)\).
A directed acyclic graph (DAG) encoding conditional independences which are compatible with Assumption~\ref{asn:identify} is displayed in Fig.~\ref{fig:hidden}.
Since the full data distribution can satisfy this assumption in various ways, we list additional examples of compatible graphs in Fig.~\ref{fig:dag-add} in the Supplementary Material.
Some differences to Fig.~\ref{fig:hidden} include \(Z\) being a cause of \(U\) (\(Z\to U\)) and \(W\) sharing unmeasured causes with \(Y\) (\(W\leftrightarrow Y\)).
However, no direct edge may exist between \(Z\) and \((A,Y,W)\) or between \(W\) and \((A,S,Z)\).

Problem-specific effect measures allow for the relaxation of the distribution-level Assumption~\ref{asn:identify}\ref{eqn:identify-zy}.
In this work, we work with the CATE transportability Assumption~\ref{asn:causal}\ref{asn:transportability}, and the target parameter is an ATE.
It suffices to assume that
\begin{multline*}
  \E(Y\mid A=1,Z,X,U,S=1)-\E(Y\mid A=0,Z,X,U,S=1)\\
  =\E(Y\mid A=1,X,U,S=1)-\E(Y\mid A=0,X,U,S=1).
\end{multline*}
That is, we allow \(Z\) and \(Y\) to be dependent conditioning on \((A,X,U)\) in the source RCT, as long as \(Z\) is not an effect modifier of \(A\) on \(Y\) on the CATE scale after adjusting for \((X,U)\).

Heuristically, we want the adjustment proxy \(W\) to emulate the effect of \(U\) in the mean outcome difference model.
Similarly, the reweighting proxy \(Z\) should take the place of \(U\) in the participation odds model.
To this end, we introduce two sets of functions involving the proxies, which are defined on the full data distribution:
\begin{align*}
  \H^{\u}&= \big\{h^{\u}(w,x)\in L_{2}(W,X;P^1):\E\{\tilde{Y}-h^{\u}(W,X)\mid X,U,S=1\}=0\big\},\\
  \Q^{\u}&= \bigg\{q^{\u}(z,x)\in L_{2}(Z,X;P^1):\E\{q^{\u}(Z,X)\mid X,U,S=1\} = \dfrac{P^{\u}(S=0\mid X,U)}{P^{\u}(S=1\mid X,U)}\bigg\},
\end{align*}
where the conditions inside the sets hold \(P^{\u}(X,U\mid S=0)\)-almost surely, and \(L_{2}(V;P^1)\) is the space of square integrable functions of \(V\) with respect to the probability measure \(P^1\).

We refer to the elements of \(\H^{\u}\) and \(\Q^{\u}\), if they exist, as outcome bridge functions and participation bridge functions, respectively.
These bridge functions recover the unobserved nuisance functions in Proposition~\ref{ppn:identification-unmeasured} after being projected onto a subspace of the full data distribution.

\begin{assumption}
  \label{asn:nonempty}
  Either \(\H^{\u}\) or \(\Q^{\u}\) is nonempty.
\end{assumption}

A sufficient condition for \(\H^{\u}\) or \(\Q^{\u}\) to be nonempty is a relevance assumption for the proxy \(W\) or \(Z\), stating that the proxy should at least be correlated with \(U\) after controlling for the baseline covariates.
If the bridge functions do not exist, measurements of the proxies will not grant more information on \(U\) than what can be inferred from adjusting for \(X\), making them ineffective in mimicking the dependence of \(U\) on \(S\) and of \(Y\) on \(U\).
See the discussions in \citet{kallus2022causal}, Examples 3 and 4.
Some examples of potential unobserved effect modifiers are the standard of care and the social determinants of health among the RCT participants.
These are often high-dimensional covariates that are not available from the experimental setting.
Nonetheless, it may be reasonable to posit that only a low-dimensional subset of these covariates strongly contributes to effect modification, such as concomitant medication and employment stability.
If this is true, then the number of proxies does not have to be large for bridge functions to exist.

On the observed data distribution, we define two sets of observed data bridge functions:
\begin{align*}
  \H &= \big\{h(w,x)\in L_{2}(W,X;P^1):\E\{\tilde{Y}-h(W,X)\mid Z,X,S=1\}=0\big\},\\
  \Q &= \bigg\{q(z,x)\in L_{2}(Z,X;P^1):\E\{q(Z,X)\mid W,X,S=1\} = \dfrac{P(S=0\mid W,X)}{P(S=1\mid W,X)}\bigg\},
\end{align*}
where the condition inside \(\H\) holds on the set \(\{(z,x):P(S=1\mid Z=z,X=x)P(S=0\mid X=x)>0\}\), and the condition inside \(\Q\) holds \(P(W,X\mid S=0)\)-almost surely.

Under the conditional independences in Assumption~\ref{asn:identify}, we can relate outcome bridge functions \(h^{\u}(w,x)\) to the reweighting proxy \(Z\) and participation bridge functions \(q^{\u}(z,x)\) to the adjustment proxy \(W\).
The existence of the bridge functions on the full data distribution implies their existence on the observed data distribution under proper proxy assumptions.

\begin{lemma}
  \label{lem:bridge}
  If Assumptions~\ref{asn:identify}\ref{eqn:identify-a}--\ref{asn:identify}\ref{eqn:identify-zy} hold, then \(\H^{\u}\subset\H\).
  If Assumptions~\ref{asn:identify}\ref{eqn:identify-zw} and \ref{asn:identify}\ref{eqn:identify-ws} hold, then \(\Q^{\u}\subset\Q\).
\end{lemma}

The proof of Lemma~\ref{lem:bridge} can be found in Supplementary Material \S\ref{sec:bridge-proof}.
The sets of observed bridge functions give rise to the following identifiability result.

\begin{proposition}[Identifiability]
  \label{ppn:identification}
  Suppose Assumptions~\ref{asn:causal}--\ref{asn:nonempty} hold.
  If \(\H\neq \emptyset\) and \(\Q\neq \emptyset\), then the target parameter \(\theta\) is identifiable in the observed data distribution \(P\).
  For any \(h\in\H\),
  \[
    \theta=\E\{h(W,X)\mid S=0\},
  \]
  and for any \(q\in\Q\),
  \[
    \theta=\frac{1}{\alpha}\E\{Sq(Z,X)\tilde{Y}\}.
  \]
\end{proposition}

The proof of Proposition~\ref{ppn:identification} can be found in Supplementary Material \S\ref{sec:identification-proof}.
If there exist solutions to both of the equations in \(\H\) and \(\Q\), then they can essentially be regarded as the unobservable bridge functions in \(\H^{\u}\) and \(\Q^{\u}\) in the identification of the target parameter.

Rather than CATE transportability, an indirect comparison can also assume mean transportability \(\E\{Y(1)\mid X,U,S=1\}=\E\{Y(1)\mid X,U,S=0\}\).
An analogous assumption to the existence of an outcome difference bridge function \(h^{\u}\in\H^{\u}\) is the existence of functions \(\tilde{h}^{\u}(w,x)\) such that \(\E\{Y-\tilde{h}^{\u}(W,X)\mid X,A=1,U,S=1\}=0\).
This is referred to as unanchored indirect comparison \citep{phillippo2018methods}.
Unanchored indirect comparison is valid only when all shifted prognostic variables, observed and unobserved, are taken into account.
From a practical point of view, the existence of bridge functions is related to the explanatory power of the proxies relative to the unobserved variables.
The set of proxies required for identifiability of the causal parameter \(\E\{Y(1)\mid S=0\}\) via the outcome bridge \(\tilde{h}^{\u}\) is potentially much larger than what is needed for the identifiability of \(\theta\) via the outcome difference bridge \(h^{\u}\).

\subsection{Connections to other proximal causal inference approaches}

\citet{ghassami2022combining} described a method for estimating long-term treatment effects exploiting the internal validity of experimental data while using proxies to account for confoundedness in observational data.
The target population is the one defined by the observational data, and the short-term outcome plays the role of outcome-inducing proxy \(W\).
The treatment bridge function \(q\) in their work is intended to emulate the inverse of the propensity score in the confounded data, which depends on unobserved confounders \(U\).
\citet{imbens2024longterm} also employed proximal causal inference in their attempt to estimate long-term treatment effect via data fusion.
The selection bridge function assumed in their work has a similar form as the participation bridge in \(\Q\), in that they both capture the variability of the unobserved variables \(U\) between two data sources.

However, the problem studied in these two articles is fundamentally different from ours, since the authors make a ``data combination'' assumption, which is Assumption 3 in \citet{ghassami2022combining} and Assumptions 3 and 10 in \citet{imbens2024longterm}.
The assumption is that the unobserved variables are independent of the RCT indicator \(S\), possibly conditioning on some baseline covariates \(X\), but we do not assume this.
In proximal indirect comparison, we wish to capture the difference in distributions of unobserved variables between the study populations through the use of proxies.
Additionally, both bridge functions in these works are assumed on the target population, whereas the bridge functions \(\H\) and \(\Q\) in proximal indirect comparison are posited on the source population.
In particular, the outcome bridge cannot be defined nor learned on the target population due to the impossibility of observing a treatment of interest.

In the appendix of their work, \citet{imbens2024longterm} provide an additional identification result, allowing unobserved variables to vary between the experimental population and the observational population.
Instead, they require that the treatment is completely randomized and that latent transportability holds on the treatment-specific means.
Unlike their setup, we avoid making these assumptions, since our bridge functions do not involve specific treatments.
Our Proposition~\ref{ppn:identification} remains valid under stratified randomization, which is common in RCTs.

\citet{ghassami2024causal} proposed a closely related approach for causal mediation analysis and front-door adjustment in the presence of hidden mediators.
For the identifiability of controlled and uncontrolled treatment-specific means, they assumed a treatment bridge function describing the difference between the unobserved variables under different interventions (their Assumption 9).
Despite the resemblance of the bridge functions in our works, they are different as we are interested in relaxing the CATE transportability for the estimation of the treatment effect with at least one treatment arm that is not present in the target population.
The unmeasured variables in indirect comparison cannot be mediators in that the CATE transportability in Assumption~\ref{asn:causal}\ref{asn:transportability} is conditional on \(U\).
In addition, the RCT indicator \(S\) is not an intervention nor exposure, so conceptually \(U\) cannot be considered as a mediator between \(S\) and \(Y\).

We make an important remark that for data fusion, it is often not necessary to observe all groups of proxies in both populations.
For example, in \citet{ghassami2022combining} the treatment-inducing proxy is only required in the observational regime.
The short-term outcome in \citet{imbens2024longterm}, which enjoys similar properties as a treatment-inducing proxy, appears also in the observational data only.
In proximal indirect comparison, no reweighting proxy \(Z\) needs to be observed in the target population.

\section{Asymptotic theory for target parameter estimation}
\label{sec:asymptotics}

\subsection{A nonparametric influence function}

Before presenting an estimator for the target parameter, we make a connection between the sampling scheme and the probability model.
In the source trial \(S_{i}=1\), the observed data is an independently and identically distributed (i.i.d.) sample \((A_{i},X_{i},W_{i},Z_{i},Y_{i})\), \(i=1,2,\dots,n_{1}\).
In the target trial \(S_{i}=0\), the observed data is an i.i.d. sample of \((X_{i},W_{i})\), \(i=n_{1}+1,n_{1}+2,\dots,n_{1}+n_{0}\).
For the asymptotic arguments, we require that the ratio \(n_{0}/n\) approach the fixed number \(\alpha\) between zero and one when \(n\) goes to infinity, where \(n=n_{0}+n_{1}\) is the total number of observations.
Then we can consider the observed data as an i.i.d. sample from \(P_{0}\), the true data generating mechanism.
In this section, we use subscript \(0\) to indicate dependence on \(P_{0}\).
We write \(\tilde{Y}_0=(2A-1)Y/e_0(A\mid X)\).
We deal with the true parameter \(\theta_{0}\), which has a causal interpretation under the assumptions in Proposition~\ref{ppn:identification}.

We study the target parameter with tools from semiparametric estimation theory under the following regularity conditions on the true data distribution \(P_{0}\).
Let the linear transformation \(T_{0}:L_{2}(W,X;P^{1}_{0})\to L_{2}(Z,X;P^{1}_{0})\) be such that \((T_{0}h)(z,x)=\E_{P_{0}}\{h(W,X)\mid Z=z,X=x,S=1\}\).
Moreover, let \(F_0\) be the linear transformation \((F_{0}g)(z,x)=\E_{P_0}[\{\tilde{Y}_0-h_{0}(W,X)\}g(Y,Z,W,A,X)\mid Z=z,X=x,S=1]\) where \(g\in L_2(Y,Z,W,A,X;P^1_0)\).

\begin{assumption}[Regularity conditions]
  \label{asn:tangent}
  \hfill
  \begin{enumerate}[label=(\roman*)]
  \item\label{eqn:boundedness} \(\E_{P_0}(\tilde{Y}_0\mid Z=z,X=x,S=1)\in L_2(Z,X;P_0^1)\), \(P_0(S=1\mid W=w,X=x)/P_0(S=0\mid W=w,X=x)\in L_2(W,X;P_0^1)\);
  \item\label{eqn:bijectivity} \(T_{0}\) is bijective;
  \item\label{eqn:score} \(\mathrm{range}(F_0)\subset L_2(Z,X;P^1_0)\);
    
  \item\label{eqn:propensity} \((Z,W)\indep_{P_{0}} A\mid (X,S=1)\) and \(e_{0}(a\mid x)\) is known.
  \end{enumerate}
\end{assumption}

Define the model
\[
  \P = \{P\in\mathcal{M}: (Z,W)\indep A\mid (X,S=1), e(a\mid x)\text{ known}, \H\neq\emptyset\},
\]
where \(\mathcal{M}\) is the set of all probability measures over \(O\).
Assumptions
\ref{asn:tangent}\ref{eqn:boundedness}--\ref{asn:tangent}\ref{eqn:score} are not necessary for proposing regular and asymptotically linear estimators for \(\theta_0\), they enable the characterization of all such estimators of \(\theta_{0}\) under \(\P\).
We do not consider the propensity score \(e_0(a\mid x)\) as a nuisance function, since the treatments in RCTs are usually administered according to a predetermined protocol.



It is clear that under Assumptions~\ref{asn:tangent}\ref{eqn:boundedness}--\ref{eqn:bijectivity}, we have \(\H_{0}=\{h_{0}\}\) and \(\Q_{0}=\{q_{0}\}\).
This observation has two implications.
First, \(\H_0\) and \(\Q_0\) are simultaneously nonempty.
Working under boundedness conditions related to the projected variance of the bridge functions, \citet{zhang2023proximal} show that this condition is necessary for the observed data functional \(\theta_0\) to be \(n^{1/2}\)-estimable.
Second, the bridge functions \(h_{0}\) and \(q_{0}\) are unique.
If other bridge functions existed, the target parameter would be a uniquely identified functional on possibly nonunique nuisance parameters \citep{zhang2023proximal,bennett2023inference}.
Although this would pose no difficulty to identification of the target parameter, it would largely hinder the study of estimators constructed using estimates of the bridge functions.
In Supplementary Material \S\ref{sec:bridge-app}, we give sufficient conditions for establishing the existence and uniqueness of the observed data bridge functions, namely completeness assumptions and more regularity conditions on \(T_{0}\) and its adjoint.

We are now in a position to present a useful characterization of the target parameter by an influence function.

\begin{proposition}
  \label{ppn:eif}
  Suppose Assumption~\ref{asn:tangent} holds.
  An influence function of the observed data target parameter \(\theta_{0}\) under the model \(\P\) is
  \[
  \phi_{0}(o) = \frac{s}{\alpha_{0}}q_{0}(z,x)\{\tilde{y}_{0}-h_{0}(w,x)\}+\frac{1-s}{\alpha_{0}}\{h_{0}(w,x)-\theta_{0}\}.
\]
\end{proposition}

The proof of Proposition~\ref{ppn:eif} can be found in Supplementary Material \S\ref{sec:eif-proof}.
In fact, we can find many influence functions of \(\theta_{0}\) under the model restrictions specified in Assumption~\ref{asn:tangent}.
We purposefully choose to present \(\phi_{0}\) over the efficient one, which involves an additional nuisance parameter \(\E_{P_{0}}(Y\mid Z=z,W=w,A=a,X=x,S=1)\).
The efficient influence function can be found in the proof of the proposition in  Supplementary Material \S\ref{sec:proof-app}.
In the sequel, we construct an estimator for \(\theta_{0}\) from \(\phi_{0}\), which is preferred because we only need to estimate the bridge functions \(h_0\) and \(q_0\).
While we do not pursue the possibility here, we remark that asymptotically efficient estimators, which attain the semiparametric efficiency bound, may be constructed from the efficient influence function.

\subsection{Doubly robust estimation}

In general, we do not have the knowledge of the bridge functions.
Therefore, we resort to two-stage data-adaptive estimation of the target parameter.
In the first stage, we use the observed data to obtain estimates of the bridge functions, which are nuisance functions to the estimation problem.
In the second stage, we plug in the estimated bridge functions to some valid estimating equation for the target parameter, and an estimate of the target parameter is obtained as the solution to the estimating equation.

Suppose \(\hat{h}\) and \(\hat{q}\) are nonparametric or semiparametric estimators intended for the true, unique bridge functions \(h_{0}\) and \(q_{0}\).
Let \(\hat{\alpha}=n_{0}/n\) be the proportion of samples from the target RCT.
Propositions~\ref{ppn:identification} and \ref{ppn:eif}, together with the form of the influence function \(\phi_{0}\), suggest a natural estimator of the target parameter \(\theta_0\) as
\[
  \hat{\theta} = \frac{1}{n}\sum_{i=1}^{n}\bigg[\frac{S_{i}}{\hat{\alpha}}\hat{q}(Z_{i},X_{i})\{\tilde{Y}_{i}-\hat{h}(W_{i},X_{i})\}+\frac{1-S_{i}}{\hat{\alpha}}\hat{h}(W_{i},X_{i})\bigg].
\]
The estimator \(\hat\theta\) solves the estimating equation based on \(\phi_0\).
We now characterize its asymptotic behavior.

\begin{assumption}[Regularity conditions]
  \label{asn:asymptotics}
  \hfill
  \begin{enumerate}[label=(\roman*)]
  \item The function class
    \[
      \mathcal{G}_{0}=\bigg\{g(o)=\frac{s}{\alpha'}q'\{\tilde{y}-h'\}+\frac{1-s}{\alpha'}h':\alpha'\in[0,1],h'\in L_{2}(W,X;P^{1}_{0}),q'\in L_{2}(Z,X;P^{1}_{0})\bigg\}
    \]
    is \(P_{0}\)-Donsker;
  \item There exists a universal constant \(M> 1\) such that \(\alpha_0\geq M^{-1}\), \(\hat{\alpha}\geq M^{-1}\), \(e_{0}\geq M^{-1}\), \(|h_{0}|\leq M\), \(|\hat{q}|\leq M\), \(P_{0}(S=1\mid W,X)\geq M^{-1}\), and \(\E_{P_{0}}(Y^{2}\mid Z,X,S=1)\leq M\).
  \end{enumerate}
\end{assumption}

\begin{theorem}
  \label{thm:asymptotics}
  Suppose Assumptions~\ref{asn:tangent} and \ref{asn:asymptotics} hold and that \(\|\hat{h}-\bar{h}\|_{P^1_0}=o_{P_0}(1)\), \(\|\hat{q}-\bar{q}\|_{P^1_0}=o_{P_0}(1)\) for some nonrandom functions \(\bar{h}(w,x)\), \(\bar{q}(z,x)\) in \(L_{2}(P^{1}_{0})\).
  Then:
  \begin{enumerate}
  \item The estimator \(\hat{\theta}\) is consistent for \(\theta_{0}\), if either \(\bar{h}=h_{0}\) or \(\bar{q}=q_{0}\).
  \item The estimator \(\hat{\theta}\) is asymptotically linear with influence function \(\phi_{0}\), if \(\bar{h}=h_{0}\), \(\bar{q}=q_{0}\), and \(\|\hat{q}-q_{0}\|_{P^{1}_{0}}\|\hat{h}-h_{0}\|_{P^{1}_{0}}=o_{P_{0}}(n^{-1/2})\).
  \end{enumerate}
\end{theorem}

The proof of Theorem~\ref{thm:asymptotics} can be found in Supplementary Material \S\ref{sec:asymptotics-proof}.
The Donsker class condition on \(\mathcal{G}_{0}\) can be relaxed by applying cross-fitting to the estimation of the bridge functions.
When the bridge functions are estimated using minimax criteria, their convergence in the \(L_{2}(P_0^1)\)-norm can be established following the arguments from \citet{kallus2022causal}.
A key assumption is the equivalence of the \(L_{2}(P_{0}^1)\)-norm of the bridge functions and that of the projected bridge functions, the latter of which is easier to bound.
For the outcome bridge function, it amounts to \(\|\hat{h}-h_{0}\|_{P^{1}_0}=O_{P_0}\{\|T_{0}(\hat{h}-h_{0})\|_{P^{1}_0}\}\).
The bijectivity of \(T_{0}\) from Assumption~\ref{asn:tangent}\ref{eqn:bijectivity} is sufficient for the norm equivalence, if \(T_{0}\) is also a bounded operator.

The estimator \(\hat{\theta}\) is doubly robust in the sense that it is consistent if either the outcome bridge function \(h_{0}\) or the participation odds bridge function \(q_{0}\) is correctly estimated.
Moreover, the estimator is asymptotically normal under the observed data model \(\P\), when both bridge functions converge sufficiently fast to the ground truth, for example, both at the subparametric \(o_{P_0}(n^{-1/4})\)-rate.
In this case, if the function class \(\mathcal{G}_0\) has a square-integrable envelope, the squared empirical \(L_{2}(P_0)\)-norm \(n^{-1}\sum_{i=1}^{n}\hat{\phi}^{2}(O_{i})\) is a consistent estimator of the asymptotic variance of the estimator, where we obtain \(\hat{\phi}\) by plugging the nuisance estimates into \(\phi\).
We show this in Supplementary Material \S\ref{sec:asymptotics-proof}.

\section{Numerical results}
\label{sec:numerical}

\subsection{Simulated data example}
\label{sec:simulation}
To illustrate the method of proximal indirect comparison in numerical studies, we posit parametric models for the bridge functions.
Note however, in general the parametric assumptions are not necessary, and the asymptotic properties of the estimator \(\hat{\theta}\) in \S\ref{sec:asymptotics} hold for nonparametric estimators for the bridge functions.
Following \citet{cui2024semiparametric}, the finite-dimensional parameters in the bridge functions are estimated by the generalized method of moments motivated by their influence functions.
The details are available in Supplemantary Material \S\ref{sec:asymp-bridge-app}.

We generate the full data \((X,U,S,SA,SZ,W,SY)\) sequentially from the distributions described below.
The baseline covariates \(X\sim \Phi[\mathrm{Normal}\{(0,0,0)^{\T},\Sigma\}]\), where \(\Phi\) is the standard normal distribution function and the covariance matrix is
\[
  \Sigma=
  \begin{pmatrix}
    1 & 0.25 & 0.25 \\
    0.25 & 1 & 0.25 \\
    0.25 & 0.25 & 1
  \end{pmatrix},
\]
are correlated and have \(\mathrm{Uniform}([0,1])\) marginals.
The rest of the variables are obtained in the following way:
\begin{align*}
  U &\sim \mathrm{Uniform}([-1,0]\times [-1,0]\times [-1,0]), \\
  S\mid (X,U) &\sim \mathrm{Bernoulli}\{\mathrm{expit}(-0.625+0.5X^{\T}1+0.5U^{\T}1)\}, \\
  A\mid (X,S=1) &\sim \mathrm{Bernoulli}(0.5), \\
  Z\mid (X,U,S=1) &\sim \mathrm{Normal}(U+X,0.25\mathrm{Id}), \\
  W\mid (X,U) &\sim \mathrm{Normal}(U+X,0.25\mathrm{Id}), \\
  Y\mid (W,A,X,U,S=1) &\sim \mathrm{Normal}(0.5-A+U^{\T}1+AU^{\T}1+X^{\T}1+W^{\T}1+AW^{\T}1,0.5^{2}).
\end{align*}
The parameters are selected so that the probability \(\alpha\) is close to \(0.65\).
In this data generating mechanism, \(U\) is an effect modifier and the target parameter is \(\theta=\E\{\E(Y\mid A=1,X,U,S=1)-\E(Y\mid A=0,X,U,S=1)\mid S=0\}\).

Let \(b(z,x)=(1,z^{\T},x^{\T})^{\T}\), \(c(w,x)=(1,w^{\T},x^{\T})^{\T}\).
As we show in Supplementary Material \S\ref{sec:simulation-app}, the underlying bridge functions are unique and have the closed forms \(h_{\eta_{0}}(w,x) = \eta_{0}^{\T}c(w,x)\) and \(q_{\xi_{0}}(z,x) = \exp\{\xi_{0}^{\T}b(z,x)\}\), where \(\eta_{0}\) and \(\xi_{0}\) are nuisance parameter vectors of appropriate dimensions. 
We compare three estimators, namely \(\hat{\theta}\) proposed in \S\ref{sec:asymptotics}, as well as
\begin{align*}
\hat{\theta}_{h} &=\frac{1}{n_{0}}\sum_{i:S_{i}=0}h_{\hat{\eta}}(W_{i},X_{i}), \\ \hat{\theta}_{q} &=\frac{1}{n_{0}}\sum_{i:S_{i}=1}q_{\hat{\xi}}(Z_{i},X_{i})\tilde{Y}_{i}.
\end{align*}
The nuisance parameter estimators with correctly specified \(h_{\eta}\) and \(q_{\xi}\) were obtained on the full sample via the generalized method of moments:
\begin{align*}
  \hat{\eta}&=\arg\underset{\eta}{\min}\;\bigg\|\frac{1}{n_{1}}\sum_{i:S_{i}=1}{b}(Z_{i},X_{i})\{\tilde{Y}_{i}-{h}_{\eta}(W_{i},X_{i})\}\bigg\|^{2}, \\
  \hat{\xi}&=\arg\underset{\xi}{\min}\;\bigg\|\frac{1}{n}\sum_{i=1}^{n}\{c(W_{i},X_{i})\}^{3}\{S_{i}q_{\xi}(Z_{i},X_{i})-(1-S_{i})\}\bigg\|^{2}.
\end{align*}
The cubic of the function \(c(w,x)\) makes sure that the estimators \(\hat{\theta}\) and \(\hat{\theta}_{q}\) are numerically distinguishable, as the true bridge function \(h_{\eta}(w,x)\) is linear.
We present details of the claim in Supplementary Material \S\ref{sec:simulation-app}.

To contrast the behavior of the estimators under model misspecifications, we considered configurations where neither \(h\) nor \(q\) was misspecified (experiment 1), where \(q\) was misspecified (experiment 2), where \(h\) was misspecified (experiment 3), and where both \(h\) and \(q\) were misspecified (experiment 4).
The misspecified models were fitted by replacing \(W\) and \(Z\) with \(|W|^{1/2}\) and \(|Z|^{1/2}\) wherever appropriate.
Summary statistics of the estimators from \(1000\) repeated samples of size \(n\in\{1000,2000\}\) are displayed in Table~\ref{tab:sim}, where the reference Monte-Carlo target parameter was calculated by static interventions of \(A\) when \(S=0\).
The standard errors of the estimators for \(\hat{\theta}_{h}\) and \(\hat{\theta}_{q}\) were obtained by plugging in the nuisance parameter estimates in the theoretical asymptotic variances shown in Supplementary Material \S\ref{sec:asymp-bridge-app}.
The estimators \(\hat{\theta}_{h}\) and \(\hat{\theta}_{q}\) showed little bias only when \(h\) and \(q\) were correctly estimated, respectively.
This was constrasted by \(\hat{\theta}\), which exhibited the double robustness property as expected.
The influence-function-based standard error for \(\hat{\theta}\) also showed robustness against model misspecification.
In Supplementary Material \S\ref{sec:simulation-app}, we present additional simulations under alternative data generating mechanisms to investigate the behavior of the proximal estimators, including invalid proxies, nonunique bridge functions, weak proxies and near violation of positivity.
In particular, we found that ridge regularization on the parameters \(\eta\) and \(\xi\) recovered valid inference for the doubly robust estimator when bridge functions were nonuniquely defined.

For comparison, we repeated the simulation for the doubly robust estimator in \citet{dahabreh2020extending}, where only \(X\) was used for standardization.
The outcome mean \(\E(Y\mid A=a,X=x,S=1)\) and the selection score \(\Pr(S=1\mid X=x)\) were fitted by ordinary least squares separately for \(a\in\{0,1\}\) and a logistic regression, respectively.
Since the outcome mean model is correctly specified, we should expect the estimator to be consistent for the estimand \(\E\{\E(Y\mid A=1,X,S=1)-\E(Y\mid A=0,X,S=1)\mid S=0\}\).
The results collected in Table~\ref{tab:sim-dahabreh} show that in the presence of unmeasured effect modifiers, the estimator is empirically biased for \(\theta\) as expected.

\begin{table}
  \caption{Simulation results of experiments 1--4.}
  \label{tab:sim}
  \footnotesize
  \centering
  {\begin{tabular}{lllS[]S[]rrr}
    \toprule
    {\(n\)} & {Experiment} & {Estimator} & {Mean} & {Bias} & {RMSE} & {SE} & {Coverage}\\
    \midrule
    \(1000\) & 1 & \(\hat{\theta}_{h}\) & -2.64 & 1.03 & 3.17 & 3.21 & 95.7\\
            &  & \(\hat{\theta}_{q}\) & -2.65 & -9.37 & 3.96 & 3.99 & 95.4\\
            &  & \(\hat{\theta}\) & -2.64 & 3.28 & 3.70 & 3.75 & 95.4\\
    \cmidrule{2-8}
            & 2 & \(\hat{\theta}_{h}\) & -2.64 & 1.03 & 3.17 & 3.21 & 95.7\\
            &  & \(\hat{\theta}_{q}\) & -2.39 & 251.28 & 3.95 & 3.16 & 86.2\\
            &  & \(\hat{\theta}\) & -2.64 & 5.69 & 3.42 & 3.08 & 92.7\\
    \cmidrule{2-8}
            & 3 & \(\hat{\theta}_{h}\) & -2.39 & 257.16 & 3.91 & 3.05 & 86.6\\
            &  & \(\hat{\theta}_{q}\) & -2.65 & -9.37 & 3.96 & 3.99 & 95.4\\
            &  & \(\hat{\theta}\) & -2.65 & -3.10 & 4.07 & 3.84 & 93.8\\
    \cmidrule{2-8}
            & 4 & \(\hat{\theta}_{h}\) & -2.39 & 257.16 & 3.91 & 3.05 & 86.6\\
            &  & \(\hat{\theta}_{q}\) & -2.39 & 251.28 & 3.95 & 3.16 & 86.2\\
            &  & \(\hat{\theta}\) & -2.39 & 251.68 & 3.93 & 3.16 & 87.1\\
    \midrule
    \(2000\) & 1 & \(\hat{\theta}_{h}\) & -2.65 & -5.83 & 2.28 & 2.25 & 94.3\\
            &  & \(\hat{\theta}_{q}\) & -2.66 & -17.61 & 2.69 & 2.68 & 95.3\\
            &  & \(\hat{\theta}\) & -2.65 & -9.46 & 2.55 & 2.55 & 94.8\\
    \cmidrule{2-8}
            & 2 & \(\hat{\theta}_{h}\) & -2.65 & -5.83 & 2.28 & 2.25 & 94.3\\
            &  & \(\hat{\theta}_{q}\) & -2.40 & 243.48 & 3.22 & 2.14 & 79.0\\
            &  & \(\hat{\theta}\) & -2.65 & -5.31 & 2.40 & 2.11 & 92.3\\
    \cmidrule{2-8}
            & 3 & \(\hat{\theta}_{h}\) & -2.40 & 249.97 & 3.25 & 2.08 & 77.3\\
            &  & \(\hat{\theta}_{q}\) & -2.66 & -17.61 & 2.69 & 2.68 & 95.3\\
            &  & \(\hat{\theta}\) & -2.66 & -14.11 & 2.76 & 2.59 & 94.1\\
    \cmidrule{2-8}
            & 4 & \(\hat{\theta}_{h}\) & -2.40 & 249.97 & 3.25 & 2.08 & 77.3\\
            &  & \(\hat{\theta}_{q}\) & -2.40 & 243.48 & 3.22 & 2.14 & 79.0\\
            &  & \(\hat{\theta}\) & -2.40 & 243.56 & 3.22 & 2.14 & 78.6\\
    \bottomrule
  \end{tabular}}

\medskip
  {Bias: Monte-Carlo bias, \(10^{-3}\); RMSE: root mean squared error, \(10^{-1}\); SE: average of standard error estimates, \(10^{-1}\); Coverage: \(95\%\) confidence interval coverage, \(\%\).
  Experiment 1: \(h\) and \(q\) correctly specified; experiment 2: \(q\) misspecified; experiment 3: \(h\) misspecified, experiment 4: \(h\) and \(q\) misspecified.}
\end{table}

\begin{table}
  \caption{Simulation results for the estimator in \citet{dahabreh2020extending}.}
  \label{tab:sim-dahabreh}
  \footnotesize\centering
 {\begin{tabular}{lS[]S[]rrr}
{\(n\)} & {Mean} & {Bias} & {RMSE} & {SE} & {Coverage}\\
1000 & -2.41 & 238.20 & 3.09 & 2.01 & 77.9\\
2000 & -2.40 & 242.27 & 2.84 & 2.02 & 84.8\\
 \end{tabular}}

\medskip
  {Bias: Monte-Carlo bias, \(10^{-3}\); RMSE: root mean squared error, \(10^{-1}\); SE: average of standard error estimates, \(10^{-1}\); Coverage: \(95\%\) confidence interval coverage, \(\%\).}
\end{table}

\subsection{Real data example}
\label{sec:real}

Proximal indirect comparison allows for treatment effect estimation via transportability in the presence of unobserved effect modifiers.
For a real data application of our method, we make use of the individual-level patient data from two global weight management RCTs, namely SCALE [clinicaltrials.gov ID NCT03552757, \cite{davies2015efficacy}] and STEP-2 [clinicaltrials.gov ID NCT01272232, \cite{davies2021semaglutide}].
While these trials are inherently longitudinal, we ignore this structure for the sole purpose of illustrating our method.
Whenever a subject deviates from the predetermined protocol at randomization, we treat the subsequent weight measurements as missing.

The active treatments are once-daily liraglutide, \(3.0\) mg or \(1.8\) mg in SCALE and once-weekly semaglutide, \(2.4\) mg or \(1.0\) mg in STEP-2, injected subcutaneously, both of which are glucagon-like pepetide-1 (GLP-1) agonists.
Both RCTs are placebo-controlled with placebo administration matched to their respective active treatments.
These superiority trials are designed to show the efficacy of semaglutide and liraglutide for weight loss among overweight or obese adults with type-2 diabetes.
However, the study populations in the RCTs can differ in practice due to the sampling of study participants.
Since the studies were conducted \(5\) to \(6\) years apart, a concern for the transportability of the treatment effect is the potential drift in social determinants of health which are unmeasured in both RCTs.
The main objective of the statistical analysis is to provide a head-to-head comparison of the treatments liraglutide versus semaglutide in the study population of STEP-2, taking into account unobserved social determinants.

The outcome \(Y\) is chosen as the percentage change from baseline (week \(0\)) to week \(44\) in body weight.
This is the timepoint closest to the end of treatment where body weight is measured in both RCTs.
In both RCTs, we imputed the body weight at week \(44\) with the last-observation-carried-forward principle.
In order to perform an anchored comparison, we make the assumption that the placebos used in these studies do not have any meaningful difference in their effect on the outcome, despite the differences in the frequency of administration and the volume of injection.
We restate the parameter \(\theta=\E\{Y(1)-Y(0)\mid S=0\}\), where \(A=1\) means liraglutide, \(3.0\) mg, \(A=0\) means placebo, and \(S=0\) indicates the STEP-2 trial.
To balance the study populations, we adjust for a set of baseline adjustment variables \(X=\{\)baseline body weight, age, sex, body-mass index, race, region, waist circumference, smoking status, duration of diabetes\(\}\).

\begin{figure}
  \centering
\includegraphics[scale=1]{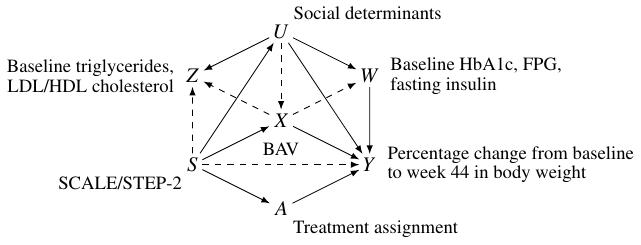}
  \caption{Hypothesized DAG of the observed and unobserved variables in the data example.
    The dashed arrows may or may not be present.
    BAV: baseline adjustment variables.}
  \label{fig:dag-data-example}
\end{figure}

Our method further requires the selection of appropriate negative controls to account for these unobserved effect modifiers.
For the adjustment proxy \(W\), we choose the percentage of glycated hemoglobin (HbA1c), the fasting plasma glucose (FPG) level and the fasting insulin level at baseline.
In a review of the impact of social determinants among type-2 diabetic patients in the United States, \citet{walker2014impact} pointed out that many studies support the link between social determinants on glycemic control measured in HbA1c.
For the reweighting proxy \(Z\), we select the baseline low-density lipoprotein (LDL) and high-density lipoprotein (HDL) cholesterol levels as well as the baseline level of triglycerides.
Cholesterol level has previously been found to be linked to health systems factors and economic development in many countries worldwide \citep{venkitachalam2012global}.
Furthermore, there is no evidence on the existence of causal pathways from the current lipid level of a person to the future body weight.
Additional assumptions on the proxies are illustrated by the causal graph in Fig.~\ref{fig:dag-data-example}.
Note that we assume the levels of HbA1c, FPG and fasting insulin differ between the study populations only because the social determinants and possibly the baseline adjustment variables are distributed differently.

\begin{table}
  \caption{Percentages of missing body weight measurements at week \(44\).}
  \label{tab:missing}
  \footnotesize\centering
  {\begin{tabular}{lcccc}
    & \multicolumn{2}{c}{SCALE} & \multicolumn{2}{c}{STEP-2} \\
    & Liraglutide \(3.0\) mg & Placebo & Semaglutide \(2.4\) mg & Placebo \\
    \(N\) & \(402\) & \(205\) & \(397\) & \(393\)  \\
    Missing (\(\%\)) & \(21.39\) & \(43.41\) & \(9.82\) & \(10.69\)
   \end{tabular}}
\end{table}

To diminish skewness, the measurements for FPG, fasting insulin, HLDL cholesterol, VLDL cholesterol and triglycerides were log-transformed.
Specifically for the estimation of the bridge functions, the numerical variables among \(X\) were transformed into an orthogonal cubic basis, and ridge regularization was applied to the linear parameters.
We compared the multiply robust proximal indirect comparison estimator to the standard doubly robust estimator proposed by \citet{dahabreh2020extending}, requiring CATE transportability to hold conditional on \(X\).
The subjects with missing measurements for \(X\), \(Z\) or \(W\) and those with no body weight measurements beyond baseline were removed from the analysis, adding up to \(41/846\) in SCALE and \(26/1210\) in STEP-2.
Table~\ref{tab:missing} shows that the percentages of missing outcomes are drastically different between the subjects randomized to the liraglutide, \(3.0\) mg arm and the placebo arm in the SCALE trial.
The difference is less pronounced in the STEP-2 trial, where missingness in outcome measurements is much less frequent.
To obtain estimates of the indirect comparison estimand \(\theta\), we further estimated the average treatment effect \(\E\{Y(-1)-Y(0)\mid S=0\}\) within STEP-2 using the standard doubly robust estimator \citep{bang2005doubly}, where \(A=-1\) stands for the once-weekly semaglutide, \(2.4\) mg treatment.
A detailed description of the estimators and postulated nuisance models can be found in Supplementary Material \S\ref{sec:analysis-app}.
Standard errors for all estimators were calculated as empirical \(L_{2}\)-norms of the corresponding influence functions.

The estimates and the corresponding \(95\%\) confidence intervals are displayed in Table~\ref{tab:ic}.
The proximal estimate and the standard estimate of \(\theta\) show a similar weight loss effect of liraglutide at week \(44\).
The proximal estimate of the indirect comparison parameter \(\gamma\) is \(-3.08\%\) versus the standard estimate of \(-3.09\%\), both indicating a stronger weight loss effect of semaglutide in the study population of the STEP-2 trial.
Note, however, that the confidence interval based on the proximal estimator is slightly wider than that based on the standard estimator.
The proposed proximal estimator can be used as a tool for sensitivity analysis to CATE transportability in the observed data, or in other words, the assumption of no shifted effect modifiers beyond \(X\).
In this data example, if the modelling assumptions hold, we may postulate that the unobserved social determinants have not altered the effect of the GLP-1 agonists.

\begin{table}
  \caption{Results from the indirect comparison analysis with SCALE and STEP-2.}
  \label{tab:ic}
  \footnotesize\centering
  {\begin{tabular}{lcccc}
   Estimand & Standard & \(95\%\)-CI & Proximal & \(95\%\)-CI \\
   \(\theta=\E\{Y(1)-Y(0)\mid S=0\}\) & \(-3.80\) & \((-4.59,-3.01)\) & \(-3.82\) & \((-4.73,-2.90)\) \\
   \(\gamma=\E\{Y(1)-Y(-1)\mid S=0\}\) & \(-3.09\) & \((-4.20,-1.98)\) & \(-3.08\) & \((-4.28,-1.87)\) \\
  \end{tabular}}

\medskip
{Standard: estimators without the use of proxies detailed in Supplementary Material \S\ref{sec:analysis-app}; proximal: proximal indirect comparison estimators; CI: confidence interval.}
\end{table}

\section{Discussion}
\label{sec:discussion}

A particular challenge for applying proximal indirect comparison is the selection of proxies in RCTs.
For example, safety measurements and vital signs, which are routinely collected, usually do not affect the outcome, but they tend to be suboptimal proxy candidates because their distributions do not vary much between populations after controlling for baseline covariates.
Data linkage would allow subjects from RCTs to be identified in the health registry, thereby providing far more potential proxies to choose from.
Besides, the collection of proxy variables can be extended before and after the running period of the RCTs.
When medical history is treated as a proxy, data linkage also helps avoid the use of self-reported data from questionnaires.

There are many interesting directions for future research.
Throughout, we have assumed the availability of individual patient data in both RCTs.
However, if only aggregate data can be obtained in one of the RCTs, the data likelihood changes and the bridge functions cannot be estimated with the same integral equations.
A possible solution follows the calibration approach \citep{josey2021transporting,josey2022calibration} to balance the moments of baseline covariates and proxies between RCTs.
We have also taken the propensity score as known in the source RCT.
In practice, one may want to estimate the propensity score and rely on a two-stage procedure for estimation of the outcome bridge function.
In this setting, we may still construct a doubly robust proximal estimator and perform asymptotic analysis following \citet{kallus2022causal} and \citet{foster2023orthogonal}.
In longitudinal studies like SCALE and STEP-2 described in \S\ref{sec:real}, subjects sometimes deviate from the treatment plan or drop out before the end of studies.
The extension of proximal indirect comparison to estimating the full compliance effect is straightforward, if one is willing to assume CATE transportability at baseline \citep{breskin2021fusion} and no unmeasured time-varying confounding within RCTs.
Treating noncompliance as a form of missingness at random, we extend our estimator to this case in Supplementary Material \S\ref{sec:missing-app}.
A more general transportability framework for observational longitudinal data under weaker causal assumptions can build on \citet{ying2023proximal}.
Finally beyond indirect comparisons, network meta-analyses may compare more than two active treatments from different studies.
In a dense network, direct evidence can often be strengthened by indirect evidence.
The formulation of proximal causal inference for data fusion is left for future work.

\section*{Acknowledgement}
The authors thank Marie Thi Dao Tran from Novo Nordisk A/S for valuable input and discussions on the choices of proxies in the clinical trials SCALE and STEP-2.

\section*{Conflict of interest}
Zehao Su is funded by a research gift from Novo Nordisk A/S to the Section of Biostatistics, University of Copenhagen.
Henrik Ravn is employed by Novo Nordisk A/S.

\section*{Supplementary material}
\label{SM}
The Supplementary Material contains proofs, details of both the simulated data example and the real data example, additional simulations, and an extension of the method to missing outcomes.

\bibliography{bibliography}

\newpage

\part{}
\begin{center}
  {\large Supplementary material for ``Proximal indirect comparison''}
\end{center}

\bigskip

\renewcommand{\theequation}{S\arabic{equation}}%
\renewcommand{\thesection}{S\arabic{section}}%
\renewcommand{\thetable}{S\arabic{table}}%
\renewcommand{\thefigure}{S\arabic{figure}}%
\renewcommand{\thetheorem}{S\arabic{theorem}}%
\renewcommand{\theproposition}{S\arabic{proposition}}%
\renewcommand{\thelemma}{S\arabic{lemma}}%
\renewcommand{\theassumption}{S\arabic{assumption}}%
\renewcommand{\theremark}{S\arabic{remark}}%
\renewcommand{\thecorollary}{S\arabic{corollary}}%
\renewcommand{\thepage}{S\arabic{page}}

\setcounter{equation}{0}
\setcounter{section}{0}
\setcounter{table}{0}
\setcounter{figure}{0}
\setcounter{theorem}{0}
\setcounter{proposition}{0}
\setcounter{lemma}{0}
\setcounter{assumption}{0}
\setcounter{remark}{0}
\setcounter{corollary}{0}
\setcounter{page}{1}

\section{Proofs}
\label{sec:proof-app}

\subsection{Proof of Proposition~\ref{ppn:identification-unmeasured}}
\label{sec:identification-unmeasured-proof}
The conditional average treatment effect is
\begin{align*}
  \MoveEqLeft \E\{Y(1)-Y(0)\mid X,U,S=0\} \\
  &= \E\{Y(1)-Y(0)\mid X,U,S=1\} \tag*{[Assumption~\ref{asn:causal}\ref{asn:transportability}]} \\
  &= \E\{Y(1)\mid A=1,X,U,S=1\}-\E\{Y(0)\mid A=0,X,U,S=1\} \tag*{[Assumption~\ref{asn:causal}\ref{asn:randomization} and \ref{asn:causal}\ref{asn:positivity}]} \\
  &= \E(Y\mid A=1,X,U,S=1)-\E(Y\mid A=0,X,U,S=1). \tag*{[Assumption~\ref{asn:causal}\ref{asn:consistency}]}
\end{align*}
Proceeding from the equation above, it is immediate that the target parameter is
\begin{align*}
  \MoveEqLeft\theta=\E\{Y(1)-Y(0)\mid S=0\} \\
  &= \E[\E\{Y(1)-Y(0)\mid X,U,S=0\}\mid S=0]\\
  &= \E\{\E(Y\mid A=1,X,U,S=1)-\E(Y\mid A=0,X,U,S=1)\mid S=0\}\\
  &= \E\bigg[\E\bigg\{\frac{(2A-1)Y}{\Pr(A\mid X,U,S=1)}\,\bigg\vert\, X,U,S=1\bigg\}\,\bigg\vert\, S=0\bigg] \\
  &= \E\bigg[\E\bigg\{\frac{(2A-1)Y}{e(A\mid X)}\,\bigg\vert\, X,U,S=1\bigg\}\,\bigg\vert\, S=0\bigg] \tag*{[Assumption~\ref{asn:causal}\ref{asn:randomization}]} \\
  &= \E\{\E(\tilde{Y}\mid X,U,S=1)\mid S=0\}.
\end{align*}
Then we show the inverse probability weighting representation via the identification formula above.
Starting from the g-formula representation, we write
\begin{align*}
  \MoveEqLeft \E\{\E(\tilde{Y}\mid X,U,S=1)\mid S=0\} \\
  &=\E\bigg[\E\bigg\{\frac{S}{\Pr(S=1\mid X,U)}\tilde{Y}\,\bigg\vert\, X,U\bigg\}\,\bigg\vert\, S=0\bigg] \\
  &=\iint \E\bigg\{\frac{S}{\Pr(S=1\mid X,U)}\tilde{Y}\,\bigg\vert\, X=x,U=u\bigg\}p(u,x\mid S=0)\d u\d x,\\
  \intertext{
  and by a simple manipulation of probability densities, showing that \(p(u,x\mid S=0)=\Pr(S=0\mid X=x,U=u)p(x,u)/\Pr(S=0)\),
  }
  &=\iint \frac{1}{\alpha}\E\bigg\{S\frac{\Pr(S=0\mid X,U)}{\Pr(S=1\mid X,U)}\tilde{Y}\,\bigg\vert\, X=x,U=u\bigg\}p(u,x)\d u\d x\\
  &=\frac{1}{\alpha}\E\bigg\{S\frac{\Pr(S=0\mid X,U)}{\Pr(S=1\mid X,U)}\tilde{Y}\bigg\}.
\end{align*}

\subsection{Proof of Lemma~\ref{lem:bridge}}
\label{sec:bridge-proof}
By Assumption~\ref{asn:identify}\ref{eqn:identify-zw}, \ref{asn:identify}\ref{eqn:identify-zy} and \ref{asn:identify}\ref{eqn:identify-a} for any \(h^{\u}\in\H^{\u}\),
\[
  \E(\tilde{Y}\mid Z,X,U,S=1) = \E\{h^{\u}(W,X)\mid Z,X,U,S=1\}.
\]
The first result of the lemma is immediate after integrating both sides of the equation with respect to the conditional density of \(U\) given \((Z,X,S=1)\).

By Assumption~\ref{asn:identify}\ref{eqn:identify-zw}, for any \(q^{\u}\in\Q^{\u}\),
\[
  \E\{q^{\u}(Z,X)\mid W,X,U,S=1\}=\E\{q^{\u}(Z,X)\mid X,U,S=1\} = \frac{\Pr(S=0\mid X,U)}{\Pr(S=1\mid X,U)}.
\]
The second result of the lemma follows by integrating both sides with respect to the density \(p(u\mid W,X,S=1)\),
\begin{align*}
  \MoveEqLeft \E\{q^{\u}(Z,X)\mid X,W,S=1\} \\
  &= \int \frac{\Pr(S=0\mid X,u)}{\Pr(S=1\mid X,u)}p(u\mid W,X,S=1)\d u \\
  &= \int \frac{p(u\mid S=0,X)\Pr(S=0\mid X)}{p(u\mid S=1,X)\Pr(S=1\mid X)}\frac{p(W\mid u,X,S=1)p(u\mid S=1,X)}{p(W\mid X,S=1)}\d u \\
  &= \frac{\Pr(S=0\mid X)}{\Pr(S=1\mid X)p(W\mid S=1,X)}\int p(W\mid u,X,S=1)p(u\mid S=0,X)\d u \\
  &= \frac{\Pr(S=0\mid X)}{\Pr(S=1\mid X)p(W\mid S=1,X)}\int p(W\mid u,X,S=0)p(u\mid S=0,X)\d u \tag*{[Assumption~\ref{asn:identify}\ref{eqn:identify-ws}]}\\
  &= \frac{\Pr(S=0\mid X)p(W\mid S=0,X)}{\Pr(S=1\mid X)p(W\mid S=1,X)} \\
  &= \frac{\Pr(S=0\mid W,X)}{\Pr(S=1\mid W,X)}.
\end{align*}

\subsection{Proof of Proposition~\ref{ppn:identification}}
\label{sec:identification-proof}
Consider the case where \(\H^{\u}\neq \emptyset\).
The g-formula identification given any \(h^{\u}\in\H^{\u}\) is
\begin{align*}
  \theta &= \E\{\E(\tilde{Y}\mid X,U,S=1)\mid S=0\} \tag{Proposition~\ref{ppn:identification-unmeasured}}\\
         &= \E[\E\{h^{\u}(W,X)\mid X,U,S=1\}\mid S=0] \\
         &= \E[\E\{h^{\u}(W,X)\mid X,U,S=0\}\mid S=0] \tag*{[Assumption~\ref{asn:identify}\ref{eqn:identify-ws}]}\\
         &= \E\{h^{\u}(W,X)\mid S=0\}, \\
         &= \frac{1}{\alpha}\E\bigg\{S\frac{\Pr(S=0\mid W,X)}{\Pr(S=1\mid W,X)}h^{\u}(W,X)\bigg\} \\
  \intertext{and for any \(q\in\Q\neq\emptyset\), we can write the parameter as}
         &= \frac{1}{\alpha}\E[S\E\{q(Z,X)\mid W,X,S=1\}h^{\u}(W,X)] \\
         &= \frac{1}{\alpha}\E\{Sq(Z,X)h^{\u}(W,X)\} \\
         &= \frac{1}{\alpha}\E[Sq(Z,X)\E\{h^{\u}(W,X)\mid Z,X,S=1\}] \\
  \intertext{which by Lemma~\ref{lem:bridge} is}
         &= \frac{1}{\alpha}\E[Sq(Z,X)\E\{h(W,X)\mid Z,X,S=1\}] \\
         &= \E\{h(W,X)\mid S=0\},
\end{align*}
for any \(h\in\H\).

Now consider the case where \(\Q^{\u}\neq\emptyset\).
The inverse-probability identification given any \(q^{\u}\in\Q^{\u}\) is
\begin{align*}
  \theta &= \frac{1}{\alpha}\E\bigg\{\frac{S\Pr(S=0\mid X,U)}{\Pr(S=1\mid X,U)}\tilde{Y}\bigg\}\tag{Proposition~\ref{ppn:identification-unmeasured}}\\
         &= \frac{1}{\alpha}\E[S\tilde{Y}\E\{q^{\u}(Z,X)\mid X,U,S=1\}] \\
         &= \frac{1}{\alpha}\E[S\tilde{Y}\E\{q^{\u}(Z,X)\mid Y,A,X,U,S=1\}] \tag*{[Assumptions~\ref{asn:identify}\ref{eqn:identify-zy} and \ref{asn:identify}\ref{eqn:identify-a}]}\\
         &= \frac{1}{\alpha}\E\{Sq^{\u}(Z,X)\tilde{Y}\},\\
         &= \frac{1}{\alpha}\E\{Sq^{\u}(Z,X)\E(\tilde{Y}\mid Z,X,S=1)\},\\
  \intertext{and for any \(h\in\H\neq\emptyset\), we can write the parameter as}
         &= \frac{1}{\alpha}\E[Sq^{\u}(Z,X)\E\{h(W,X)\mid Z,X,S=1\}]\\
         &= \frac{1}{\alpha}\E\{Sq^{\u}(Z,X)h(W,X)\}\\
         &= \frac{1}{\alpha}\E[S\E\{q^{\u}(Z,X)\mid W,X,S=1\}h(W,X)],\\
  \intertext{which by Lemma~\ref{lem:bridge} is}
         &= \frac{1}{\alpha}\E[S\E\{q(Z,X)\mid W,X,S=1\}h(W,X)]\\
         &= \frac{1}{\alpha}\E\{Sq(Z,X)\tilde{Y}\},
\end{align*}
for any \(q\in\Q\).

Therefore, the parameter is identified when either (i) \(\H^{\u}\) (hence \(\H\)) and \(\Q\) are nonempty or (ii) \(\Q^{\u}\) (hence \(\Q\)) and \(\H\) are nonempty.
This is equivalent to the statement in the proposition.


\subsection{Proof of Proposition~\ref{ppn:eif}}
\label{sec:eif-proof}
We state a useful theorem in functional analysis.
The following is an adapted version of Theorem 1.3.1 from \citetsuppmat{kesavan2022nonlinear}.
\begin{theorem}[Implicit function]
  \label{thm:implicit}
  Let \(\mathcal{B}_{1}\), \(\mathcal{B}_{2}\), and \(\mathcal{B}\) be Banach spaces and let \(\Omega\in \mathcal{B}_{1}\times \mathcal{B}_{2}\) be an open subset.
  Let \(G:\Omega\to \mathcal{B}\) be a mapping such that:
  \begin{enumerate}[label=(\roman*)]
  \item \(G\) is continuous on \(\Omega\);
  \item For every \((b_1,b_2)\in\Omega\), \((\partial/\partial b_2)G(b_1,b_2)\) exists and is continuous on \(\Omega\);
  \item \(G(c_1,c_2)=0\), \((\partial/\partial b_2)G(b_1,b_2)\vert_{b_1=c_1,b_2=c_2}\) is bijective.
  \end{enumerate}
  Then there exists an open neighborhood \(\Omega_1\times\Omega_2\subset \Omega\) of \(c_1,c_2\) such that for each \(b_1\in \Omega_1\), there exists a unique, continuous function \(\rho:\Omega_1\to\Omega_2\) satisfying \(G\{b_1,\rho(b_1)\}=0\).
  Moreover, if \(G\) is differentiable at \((c_1,c_2)\), then \(\rho\) is differentiable at \(c_1\) with derivative
  \[
    \frac{\d}{\d b_1}\rho(b_1)\bigg\vert_{b_1=c_1}=-\bigg\{\frac{\partial}{\partial b_2}G(b_1,b_2)\bigg\vert_{b_1=c_1,b_2=c_2}\bigg\}^{-1}\frac{\partial}{\partial b_1}G(b_1,b_2)\bigg\vert_{b_1=c_1,b_2=c_2}
  \]
\end{theorem}

Without loss of generality, assume \(P_0\) has a density \(p_0\) with respect to a dominating measure \(\nu\).
The density \(p_0\) can be factorized as
\begin{multline*}
  p_{0}(o) = \big\{p_{0}(y\mid a,z,w,x,S=1)e_{0}(a\mid x)p_{0}(w\mid z,x,S=1)p_{0}(z\mid x,S=1)\big\}^{s}\\
  p_{0}(w\mid x,S=0)^{(1-s)}p_0(x,s),
\end{multline*}
due to the fact that \((Z,W)\indep_{P_0} A\mid (X,S=1)\).
Let \(L_{2}^{0}(P_{0})\) denote the Hilbert space of \(L_{2}(P_0)\) functions with zero mean under \(P_0\).
The tangent space \(\dot{\P}_{0}\) at \(P_0\in\P\) is the subset of \(L_{2}^{0}(P_0)\) which is the linear closure of the score functions of parametric submodels in \(\P\) that contain \(P_0\).
We claim that the tangent space is \(\dot{\P}_{0}=L_{2}^0(P_0)\setminus\Lambda\), where \(\Lambda=\{sc(z,w,x)\{a-e_0(1\mid x)\}:c(z,w,x)\in L_{2}(P^1_0)\}\).
Note that \(\dot{\P}_{0}\) is maximal.

In order to verify the claim, we find a dense subset of \(\dot{\P}_{0}\) by constructing appropriate parametric submodels.
Let \(\kappa(x)=2\{1+\exp(-2x)\}^{-1}\), so that \(\kappa(0)=1\), and \((\d\kappa/\d x)(0)=1\).
Consider the probability measure \(P_{\varepsilon}\) with density \(p_{\varepsilon}(o)\) with respect to \(\nu\) factorizable as
\begin{multline*}
  p_{\varepsilon}(o) = \{p_{\varepsilon}(y\mid a,z,w,x,S=1)e_{0}(a\mid x)p_{\varepsilon}(w\mid z,x,S=1)p_{\varepsilon}(z\mid x,S=1)\}^{s}\\
  p_{\varepsilon}(w\mid x,S=0)^{(1-s)}p_{\varepsilon}(x,s),
\end{multline*}
where
\begin{align*}
  p_{\varepsilon}(y\mid a,z,w,x,S=1) &= \frac{p_{0}(y\mid a,z,w,x,S=1)\kappa\{\varepsilon\s(y,a,z,w,x)\}}{\int p_{0}(y\mid a,z,w,x,S=1)\kappa\{\varepsilon\s(y,a,z,w,x)\}\d\nu(y)},\\
  p_{\varepsilon}(w\mid z,x,S=1) &= \frac{p_{0}(w\mid z,x,S=1)\kappa\{\varepsilon\s(w,z,x)\}}{\int p_{0}(w\mid z,x,S=1)\kappa\{\varepsilon\s(w,z,x)\}\d\nu(w)}, \\
  p_{\varepsilon}(z\mid x,S=1) &= \frac{p_0(z\mid x,S=1)\kappa\{\varepsilon\s(z, x)\}}{\int p_0(z\mid x,S=1)\kappa\{\varepsilon\s(z,x)\}\d\nu(z)}, \\
  p_{\varepsilon}(w\mid x,S=0) &= \frac{p_0(w\mid x,S=0)\kappa\{\varepsilon\s(w,x)\}}{\int p_0(w\mid x,S=0)\kappa\{\varepsilon\s(w, x)\d\nu(w)}, \\
  p_{\varepsilon}(x,s) &= \frac{p_0(x,s)\kappa\{\varepsilon\s(x,s)\}}{\int p_0(x,s)\kappa\{\varepsilon\s(x,s)\}\d\nu(x,s)},
\end{align*}
such that
\begin{align*}
  \E_{P_0}\{\s(Y, Z,W,A,X)\mid Z,W,A,X,S=1\} &= 0, \\
  \E_{P_0}\{\s(W, Z,X)\mid Z,X,S=1\} &= 0, \\
  \E_{P_0}\{\s(Z, X)\mid X,S=1\} &= 0, \\
  \E_{P_0}\{\s(W, X)\mid X,S=0\} &= 0, \\
  \E_{P_0}\{\s(X,S)\} &= 0,
\end{align*}
and that \(s\s(y, z,w,a,x)\), \(s\s(w, z,x)\), \(s\s(z, x)\), \((1-s)\s(w, x)\) and \(\s(x,s)\) are \(L_2^0(P_0)\)-functions.
By construction, \(P_{\varepsilon}\vert_{\varepsilon=0}=P_0\).
We can find an open neighborhood \(\Gamma\) around zero such that \(\{P_{\varepsilon}:\varepsilon\in\Gamma\}\) is a one-dimensional curve parametrized by \(\varepsilon\).
The propensity score \(e_0(a\mid x)\) is left out of the parameterization, since it is assumed to be known.

We now show how \(\{P_{\varepsilon}:\varepsilon\in\Gamma\}\) can be made into a regular parametric submodel.
We invoke Theorem~\ref{thm:implicit}.
Let \(\mathcal{B}_1=\Gamma\), \(\mathcal{B}_2=L_2(W,X;P^1_0)\), \(\mathcal{B}=L_2(Z,X;P^1_0)\), \(c_1=0\), and \(c_2=h_0\).
Then the mapping
\[
  G(\varepsilon,h)=\E_{P_{\varepsilon}}\{\tilde{Y}_{0}-h(W,X)\mid Z=z,X=x,S=1\}
\]
fulfils the conditions in Theorem~\ref{thm:implicit}, with \(G(0,h_0)=0\) from the definition of \(h_0\), and \((\partial/\partial h)G(\varepsilon,h)=-T_{\varepsilon}\),
where \((T_{\varepsilon}b_{2})(z,x)=\E_{P_\varepsilon}\{b_2(W,X)\mid Z=z,X=x,S=1\}\), and \((\partial/\partial h)G(\varepsilon,h)\vert_{\varepsilon=0,h=h_0}=-T_{\varepsilon}\vert_{\varepsilon=0}=-T_0\), which is bijective by Assumption~\ref{asn:tangent}\ref{eqn:bijectivity}.
It follows from Theorem~\ref{thm:implicit} that there exists a unique, continuous function \(h_{\varepsilon}(w,x)\) on an open subset \(\tilde{\Gamma}\subset\Gamma\) such that
\[
  \E_{P_\varepsilon}[\{\tilde{Y}_0-h_\varepsilon(W,X)\}\mid Z=z,X=x,S=1]=0.
\]
Therefore, \(\H_\varepsilon=\{h_\varepsilon\}\) is nonempty, which shows that \(\{P_\varepsilon:\varepsilon\in\tilde{\Gamma}\}\) is a submodel in \(\P\).

Furthermore, since
\begin{multline*}
  \frac{\partial}{\partial \varepsilon}G(\varepsilon,h)\bigg\vert_{\varepsilon=0,h=h_0}=\E_{P_0}[\{\tilde{Y}_0-h_{0}(W,X)\}\\
  \{\s(Y, Z,W,A,X)+\s(W, Z,X)\}\mid Z=z,X=x,S=1]
\end{multline*}
exists and that its range is contained in \(L_2(Z,X;P^1_0)\) by Assumption~\ref{asn:tangent}\ref{eqn:score}, the function \(h_\varepsilon\) is differentiable at \(\varepsilon=0\) with derivative
\begin{multline}
  \label{eqn:tangent-h}
  \frac{\partial}{\partial\varepsilon}h_\varepsilon(w,x)=
  \big(T_0^{-1}\E_{P_0}[\{\tilde{Y}_0-h_{0}(W,X)\}
  \\
  \{\s(Y, Z,W,A,X)+\s(W, Z,X)\}\mid Z=z,X=x,S=1]\big)(w,x)
\end{multline}
The submodel \(P_\varepsilon(\s)\) constructed in this fashion has score
\[
  \s(o)=s\{\s(y, z,w,a,x)+\s(w, z,x)+\s(z, x)\}+(1-s)\s(w, x)+\s(x,s),
\]
where the dependence on \(\s\) is written out.
The union of the tangent sets of the submodels \(\{P_\varepsilon(g):\varepsilon\in\tilde{\Gamma}\}\) obtained by varying \(g\) is
\begin{align*}
  \dot{\P}_0 &= \big\{\s(o)=s\{\s(y, z,w,a,x)+\s(w, z,x)+\s(z, x)\}+(1-s)\s(w, x)+\s(x,s): \\
             &\qquad \E_{P_0}\{\s(Y, Z,W,A,X)\mid Z,W,A,X,S=1\} = 0, \E\{\s(W, Z,X)\mid Z,X,S=1\} = 0, \\
             &\qquad \E_{P_0}\{\s(Z, X)\mid X,S=1\} = 0, \E_{P_0}\{\s(W, X)\mid X,S=0\} = 0, E_{P_0}\{\s(X,S)\} = 0,\\
             &\qquad s\s(y, z,w,a,x), s\s(w, z,x), s\s(z, x), (1-s)\s(w, x), \s(x,s) \in L_2^0(P_0)\big\}.
\end{align*}
Any function in \(L_{2}^{0}(P_0)\setminus\Lambda\) can be orthogonalized by successive projections, and each of the spaces is a closed subspace of \(L_{2}^{0}(P_0)\) corresponding to the individual functions comprising \(\s(o)\).
Therefore, \(\dot{\P}_0=L_{2}^{0}(P_0)\setminus\Lambda\), and since this tangent set is maximal, it must be the tangent space of the model \(\P\) at \(P_0\).

The target parameter is \(\theta_0=\E_{P_0}\{h_{0}(W,X)\mid S=0\}\).
The Gateaux derivative of \(\theta_\varepsilon=\E_{P_\varepsilon}\{h_{\varepsilon}(W,X)\mid S=0\}\) at \(\varepsilon=0\) along the submodel \(\{P_{\varepsilon}:\varepsilon\in\tilde{\Gamma}\}\) is
\begin{align}
  \nonumber \frac{\d}{\d\varepsilon}\theta_\varepsilon\bigg\vert_{\varepsilon=0}\ &= \frac{\partial}{\partial\varepsilon}\iint h_{\varepsilon}(w,x)p_{\varepsilon}(w,x\mid S=0)\d\nu(w,x)\bigg\vert_{\varepsilon=0} \\
                                                                                  &= \E_{P_0}\bigg\{\frac{\partial}{\partial\varepsilon}h_{\varepsilon}(W,X)\bigg\vert_{\varepsilon=0}\,\bigg\vert\, S=0\bigg\} + \iint h_{0}(w,x)\frac{\partial}{\partial\varepsilon}p_{\varepsilon}(w,x\mid S=0)\bigg\vert_{\varepsilon=0}\d\nu(w,x).\label{eqn:deriv}
\end{align}
We now study the two terms separately.
The first term in \eqref{eqn:deriv} is
\begin{flalign*}
  \MoveEqLeft\frac{1}{\alpha_0}\E_{P_0}\bigg\{S\frac{P_0(S=0\mid W,X)}{P_0(S=1\mid W,X)}\frac{\partial}{\partial\varepsilon}h_{\varepsilon}(W,X)\bigg\vert_{\varepsilon=0}\bigg\} \\
  \intertext{which, after substituting the identification equation of \(q_0\), is}
  &= \frac{1}{\alpha_0}\E_{P_0}\bigg[S\E_{P_0}\{q_{0}(Z,X)\mid W,X,S=1\}\frac{\partial}{\partial\varepsilon}h_{\varepsilon}(W,X)\bigg\vert_{\varepsilon=0}\bigg] \\
  &= \frac{1}{\alpha_0}\E_{P_0}\bigg[Sq_{0}(Z,X)\E_{P_0}\bigg\{\frac{\partial}{\partial\varepsilon}h_{\varepsilon}(W,X)\bigg\vert_{\varepsilon=0}\,\bigg\vert\, Z,X,S=1\bigg\}\bigg].\\
  \intertext{Inserting the conditional expectation of the derivative from \eqref{eqn:tangent-h}, we develop the term further as} 
  &= \frac{1}{\alpha_0}\E_{P_0}\big(Sq_{0}(Z,X)\E_{P_0}[\{\tilde{Y}_0-h_{0}(W,X)\}\{\s(Y, Z,W,A,X)+\s(W, Z,X)\}\mid Z,X,S=1]\big) \\
  &= \frac{1}{\alpha_0}\E_{P_0}[Sq_{0}(Z,X)\{\tilde{Y}_0-h_{0}(W,X)\} &\\
  && \mathllap{[S\{\s(Y, Z,W,A,X)+\s(W, Z,X)+\s(Z, X)\}+(1-S)\s(W, X)+\s(X,S)]\bigg]} \\
  &= \E_{P_0}\bigg[\frac{S}{\alpha_0}q_{0}(Z,X)\{\tilde{Y}_0-h_{0}(W,X)\}\s(O)\bigg].
\end{flalign*}
In the second to last step we added the scores \(S\s(Z, X)\) and \(\s(X,S)\), which is valid because their products with the leading factor all have zero mean due to the identification equation of \(h_{0}\).
The score \((1-S)\s(W, X)\) was added, which is allowed, since the factor \(S\) renders their product zero.

The second term in display \eqref{eqn:deriv} is
\begin{align*}
  \MoveEqLeft\frac{1}{\alpha_0}\iint \sum_{s\in\{0,1\}}(1-s)h_{0}(w,x)\frac{\partial}{\partial\varepsilon}\{p_{\varepsilon}(w\mid x,S=0)p_{\varepsilon}(s,x)\}\bigg\vert_{\varepsilon=0}\d\nu(w,x) \\
  \MoveEqLeft[1]- \frac{1}{\alpha_0^{2}}\iint \sum_{s\in\{0,1\}}(1-s)\frac{\partial}{\partial\varepsilon}\{p_{\varepsilon}(w\mid x,S=0)p_{\varepsilon}(s,x)\}\bigg\vert_{\varepsilon=0}\d\nu(w,x)\E_{P_0}\{(1-S)h_{0}(W,X)\} \\
  &= \frac{1}{\alpha_0}\iint \sum_{s\in\{0,1\}}(1-s)h_{0}(w,x)\{\s(w, x)+\s(s,x)\}p_0(w\mid x,S=0)p_0(x,s)\d\nu(w,x) \\
  &\hphantom{=}\quad - \frac{1}{\alpha_0}\iint \sum_{s\in\{0,1\}}(1-s)\theta_0\{\s(w,x)+\s(s,x)\}p_0(w\mid x,S=0)p_0(x,s)\d\nu(w,x) \\
  &= \E_{P_0}\bigg[\frac{1-S}{\alpha_0}\{h_{0}(W,X)-\theta_0\}\{(1-S)\s(W, X)+\s(X,S)\}\bigg] \\
  &= \E_{P_0}\bigg[\frac{1-S}{\alpha_0}\{h_{0}(W,X)-\theta_0\}&\\
  &\hphantom{= \E_{P_0}\bigg[}\quad [S\{\s(Y, Z,W,A,X)+\s(W, Z,X)+\s(Z, X)\}+(1-S)\s(W, X)+\s(X,S)]\bigg] \\
  &= \E_{P_0}\bigg[\frac{1-S}{\alpha_0}\{h_{0}(W,X)-\theta_0\}\s(O)\bigg].
\end{align*}
The scores \(S\s(Y\mid Z,W,A,X)\), \(S\s(W\mid Z,X)\) and \(S\s(Z\mid X)\) were added in the second to last step, which is allowed because the factor \((1-S)\) renders their products zero.
Collecting the two results above, we have that
\[
  \frac{\d}{\d\varepsilon}\theta_\varepsilon\bigg\vert_{\varepsilon=0} = \E_{P_0}\bigg(\bigg[\frac{S}{\alpha_0}q_{0}(Z,X)\{\tilde{Y}_0-h_{0}(W,X)\}+\frac{1-S}{\alpha_0}\{h_{0}(W,X)-\theta_0\}\bigg]\s(O)\bigg),
\]
which shows that \(\phi_{0}(o)\), the factor next to \(\s(o)\), is an influence function of the parameter \(\theta_0\) at \(P_0\in\P\).

Next, we will show the efficient influence function of \(\theta_0\).
Define the function
\[
  \varphi_{0}(o) = \phi_{0}(o)-\frac{s}{\alpha_0}q_{0}(z,x)\E_{P_0}\bigg[\frac{Y}{\{e_0(A\mid X)\}^{2}}\biggm\vert Z=z,W=w,X=x,S=1\bigg]\{a-e_0(1\mid x)\}.
\]
It is an influence function of \(\theta_0\), because the term after \(\phi_{0}(o)\) is an element of \(\Lambda\), and thus \((\d/\d\varepsilon)\theta_{\varepsilon}\vert_{\varepsilon=0}=\E_{P_0}\{\varphi_{0}(O)\s(O)\}\) for all \(\s\in\dot{\P}_0\).
The function expands as
\begin{flalign*}
  \varphi_{0}(o) &= \frac{s}{\alpha_0}q_{0}(z,x)\frac{(2a-1)}{e_0(a\mid x)}\{y-\E_{P_0}(Y\mid Z=z,W=w,A=a,X=x,S=1)\}\\
                 &\hphantom{=}\quad -\frac{s}{\alpha_0}q_{0}(z,x)h_{0}(w,x)+\frac{1-s}{\alpha_0}\{h_{0}(w,x)-\theta_0\}\\
                 &\hphantom{=}\quad + \frac{s}{\alpha_0}q_{0}(z,x)\frac{(2a-1)}{e_0(a\mid x)}\E_{P_0}(Y\mid Z=z,W=w,A=a,X=x,S=1) \\
                 &\hphantom{=}\quad -\frac{s}{\alpha_0}q_{0}(z,x)\sum_{a'\in\{0,1\}}\frac{\E_{P_0}(Y\mid Z=z,W=w,A=a',X=x,S=1)}{e_0(a'\mid x)}\{a-e_0(1\mid x)\}\\
                 &= \frac{s}{\alpha_0}q_{0}(z,x)\frac{(2a-1)}{e_0(a\mid x)}\{y-\E_{P_0}(Y\mid Z=z,W=w,A=a,X=x,S=1)\}\\
                 &\hphantom{=}\qquad -\frac{s}{\alpha_0}q_{0}(z,x)h_{0}(w,x)+\frac{1-s}{\alpha_0}\{h_{0}(w,x)-\theta_0\}\\
                 &\hphantom{=}\qquad +\frac{s}{\alpha_0}q_{0}(z,x)\E_{P_0}(\tilde{Y}_0\mid Z=z,W=w,X=x,S=1).
\end{flalign*}

To conclude the proof, we check that the function \(\varphi_{0}(o)\) is indeed an element of \(\dot{\P}_0\).
Consider the decomposition that
\[
  \varphi_{0}(o) = s\s^{*}(y, z,w,a,x) + s\s^{*}(w, z,x) + (1-s)\s^{*}(w, x) + \s^{*}(x,s),
\]
where
\begin{align*}
  \s^{*}(y, z,w,a,x) &= \frac{1}{\alpha_0}q_{0}(z,x)\frac{(2a-1)}{e_0(a\mid x)}\{y-\E_{P_0}(Y\mid Z=z,W=w,A=a,X=x,S=1)\}, \\
  \s^{*}(w, z,x) &= \frac{1}{\alpha_0}q_{0}(z,x)\{\E_{P_0}(\tilde{Y}_0\mid Z=z,W=w,X=x,S=1)-h_{0}(w,x)\}, \\
  \s^{*}(w, x) &= \frac{1}{\alpha_0}[h_{0}(w,x)-\E_{P_0}\{h_{0}(w,x)\mid X=x,S=0\}],\\
  \s^{*}(x,s) &= \frac{1-s}{\alpha_0}[\E_{P_0}\{h_{0}(W,X)\mid X=x,S=0\} - \theta_0].
\end{align*}
We need to check that \(\E_{P_0}\{\s^{*}(Y, Z,W,A,X)\mid Z,W,A,X,S=1\} = 0\), \(\E_{P_0}\{\s^{*}(W, Z,X)\mid Z,X,S=1\}=0\), \(\E_{P_0}\{\s^{*}(W, X)\mid X,S=0\} = 0\), as well as \(\E_{P_0}\{\s^{*}(X,S)\} = 0\), all of which hold true by the definition of \(h_{0}\) and \(\theta_0\).
Therefore, the influence function \(\varphi_{0}(o)\in\dot{\P}_{0}\) is the efficient influence function of \(\theta_0\) at \(P_0\in\P\).

\subsection{Proof of Theorem~\ref{thm:asymptotics}}
\label{sec:asymptotics-proof}
In the first part of this section, we prove Theorem~\ref{thm:asymptotics}.
In the second part, we show that the estimator of the asymptotic variance in the main text is consistent.

Let
\begin{align*}
  \ell_0(o)&=\frac{s}{\alpha_0}q_{0}\{\tilde{y}_0-h_{0}\}+\frac{1-s}{\alpha_0}h_{0},\\
  \hat{\ell}(o)&=\frac{s}{\hat{\alpha}}\hat{q}\{\tilde{y}_0-\hat{h}\}+\frac{1-s}{\hat{\alpha}}\hat{h}.
\end{align*}
Then \(\ell_0\), \(\hat{\ell}\) belong to the \(P_0\)-Donsker class \(\mathcal{G}_{0}\), which is also \(P_0\)-Glivenko-Cantelli.

\begin{proof}[of Theorem~\ref{thm:asymptotics}]
  We first show consistency.
  Consider the difference
  \begin{equation}
    \label{eqn:diff}
    \hat{\theta} - \theta_0 =(P_{n}-P_0)\hat{\ell} + \bigg(P_0\hat{\ell}-\frac{\alpha_0}{\hat{\alpha}}\theta_0\bigg)-\frac{\hat{\alpha}-\alpha_0}{\hat{\alpha}}\theta_0.
  \end{equation}
  The absolute value of the first term of \eqref{eqn:diff} is bounded by \(\sup_{g\in\mathcal{G}_{0}}|(P_{n}-P_0)g|\), which converges in probability to zero by the uniform law of large numbers applied to the \(P_0\)-Glivenko-Cantelli class \(\mathcal{G}_{0}\).
  The absolute value of the second term of \eqref{eqn:diff} is
  \begin{align}
    \bigg|P_0\hat{\ell}-\frac{\alpha_0}{\hat{\alpha}}\theta_0\bigg| &= \bigg|\frac{1}{\hat{\alpha}}P_0\{S\hat{q}(\tilde{Y}_0-\hat{h})+(1-S)\hat{h}\}-\frac{\alpha_0}{\hat{\alpha}}\theta_0\bigg| \nonumber\\
                                                                    &= \bigg|\frac{1}{\hat{\alpha}}P_0\{S\hat{q}(h_{0}-\hat{h})+Sq_{0}\hat{h}\}-\frac{\alpha_0}{\hat{\alpha}}\theta_0\bigg| \nonumber\\
                                                                    &= \bigg|\frac{1}{\hat{\alpha}}P_0\{S(\hat{q}-q_{0})(h_{0}-\hat{h})+Sq_{0}h_{0}\}-\frac{\alpha_0}{\hat{\alpha}}\theta_0\bigg| \nonumber\\
                                                                    &= \bigg|\frac{1}{\hat{\alpha}}P_0\{S(\hat{q}-q_{0})(h_{0}-\hat{h})\}\bigg| \nonumber\\
                                                                    &\leq M\|\hat{q}-q_{0}\|_{P_{0}^1}\|\hat{h}-h_{0}\|_{P^1_{0}}=o_{P_0}(1)\label{eqn:product-error}.
  \end{align}
  The fourth step above is due to the observation
  \begin{multline*}
    \E_{P_0}\{Sq_{0}(Z,X)h_{0}(W,X)\}=\E_{P_0}[Sq_{0}(Z,X)\E_{P_0}\{h_{0}(W,X)\mid Z,X,S=1\}]\\
    =\E_{P_0}\{Sq_0(Z,X)\E_{P_0}(\tilde{Y}_0\mid Z,X,S=1)\}=\E_{P_0}\{Sq_0(Z,X)\tilde{Y}_0\}=\alpha_0\theta_0.
  \end{multline*}
  The absolute value of the third term of \eqref{eqn:diff} converges in probability to zero by the trivial consistency of \(\hat{\alpha}\) and Slutsky's theorem.
  The triangle inequality shows \(\hat{\theta}\overset{\mathrm{P}}{\to}\theta_0\).
  
  We now show asymptotic linearity.
  Working under the additional assumption  \(\|\hat{q}-q_{0}\|_{P_{0}^1}\|\hat{h}-h_{0}\|_{P_{0}^1}=o_{P_0}(n^{-1/2})\), we further express the difference \eqref{eqn:diff} as
  \begin{align}
    \hat{\theta} - \theta_0 &= (P_{n}-P_0)(\hat{\ell}-\ell_0) + P_{n}\ell_0 - \theta_0 + \bigg(P_0\hat{\ell}-\frac{\alpha_0}{\hat{\alpha}}\theta_0\bigg)-\frac{\hat{\alpha}-\alpha_0}{\hat{\alpha}}\theta_0 \nonumber\\
                            &= P_{n}\bigg(\phi_{0}+\frac{1-S}{\alpha_0}\theta_0\bigg)+ (P_{n}-P_0)(\hat{\ell}-\ell_0)- \frac{2\hat{\alpha}-\alpha_0}{\hat{\alpha}}\theta_0+\bigg(P_0\hat{\ell}-\frac{\alpha_0}{\hat{\alpha}}\theta_0\bigg) \nonumber\\
                            &= P_{n}\phi_{0}+ (P_{n}-P_0)(\hat{\ell}-\ell_0)+\frac{(\hat{\alpha}-\alpha_0)^{2}}{\alpha_0\hat{\alpha}}\theta_0+o_{P_0}(n^{-1/2}). \label{eqn:diff-2}
  \end{align}
  For the last equality, we use the bound from \eqref{eqn:product-error} but with \(\|\hat{q}-q_{0}\|_{P_{0}^1}\|\hat{h}-h_{0}\|_{P_{0}^1}=o_{P_0}(n^{-1/2})\).
   Applying the central limit theorem to \(\hat{\alpha}\) and then Slutsky's theorem, the third term of \eqref{eqn:diff-2} is \(O_{P_0}(n^{-1})=o_{P_0}(n^{-1/2})\).
  The second term of \eqref{eqn:diff-2} is an empirical process term of order \(o_{P_0}(n^{-1/2})\) if \(\|\hat{\ell}-\ell_0\|_{P_0}=o_{P_0}(1)\), since \(\mathcal{G}_{0}\) is \(P_0\)-Donsker.
  To conclude the proof, we show that \(\|\hat{\ell}-\ell_0\|_{P_0}\) indeed converges in probability to zero under \(\|\hat{h}-h_{0}\|_{P_{0}^1}=o_{P_0^1}(1)\), \(\|\hat{q}-q_{0}\|_{P_{0}^1}=o_{P_0^1}(1)\) and the boundedness conditions.
  We have \(\|\hat{q}\|_{P^{1}_0}=O_{P_0}(1)\) because it is bounded, and \(\|\hat{h}\|_{P_0^1}\leq \|\hat{h}-h_{0}\|_{P_0^1}+\|h_{0}\|_{P_0^1}=o_{P_0}(1)+O_{P_0}(1)=O_{P_0}(1)\).
  Furthermore, by the boundedness of \(\hat{q}\), the terms \(\|\hat{q}(\hat{h}-h_{0})\|_{P_0^1}\) and \(\|\hat{q}h_{0}\|_{P_0^1}\) are also bounded in probability.
  The \(L_{2}(P_0)\)-norm of the plug-in function \(\hat{\ell}\) is
  \begin{align*}
    \|\hat{\ell}\|_{P_0} &\leq M \big\{\|S\hat{q}\tilde{Y}_0\|_{P_0}+\|S\hat{q}(\hat{h}-h_{0})\|_{P_0}+\|S\hat{q}h_{0}\|_{P_0}+\|(1-S)\hat{h}\|_{P_0}\big\} \\
                         &\leq M^{3}\{\E_{P_0}(SY^{2})\}^{1/2}+M\bigg[\E_{P_0}\bigg\{S\frac{P_0(S=0\mid W,X)}{P_0(S=1\mid W,X)}\hat{h}^{2}\bigg\}\bigg]^{1/2}+ O_{P_0}(1) \\
                         &\leq M^{7/2} + M^{3/2}\|\hat{h}\|_{P_0^1} + O_{P_0}(1),
  \end{align*}
  which is indeed bounded in probability.
  The \(L_{2}(P_0)\)-distance between the plugin and the true function is
  \begin{align*}
    \|\hat{\ell}-\ell_0\|_{P_0}&\leq M\big\{\|S(\hat{q}-q_{0})\tilde{Y}_0\|_{P} + \|S(\hat{q}\hat{h}-q_{0}h_{0})\|_{P_0}+\|(1-S)(\hat{h}-h_{0})\|_{P_0}+|\hat{\alpha}-\alpha_0|\|\hat{\ell}\|_{P_0}\big\},\\
    \intertext{and by similar arguments above, we bound the distance by}
                               &\leq M^{2}\|\hat{q}-q_{0}\|_{P_0^1} + M\|(\hat{q}-q_{0})h_{0}\|_{P_0^1} + M\|\hat{q}(\hat{h}-h_{0})\|_{P_0^1}  \\
                               &\hphantom{\leq}\quad + M^{1/2}\|\hat{h}-h_{0}\|_{P_0^1} + M|\hat{\alpha}-\alpha_0|O_{P_0}(1) \\
                               &\leq 2M^{2}\|\hat{q}-q_{0}\|_{P_0^1} + (M^{2}+M^{1/2})\|\hat{h}-h_{0}\|_{P_0^1} + o_{P_0}(1) = o_{P_0}(1).
  \end{align*}
  This shows \(\hat{\theta}-\theta_0=P_{n}\phi_0+o_{P_0}(n^{-1/2})\).
\end{proof}

\begin{corollary}
  Suppose the conditions for asymptotic linearity of \(\hat\theta\) in Theorem~\ref{thm:asymptotics} hold.
  Furthermore, \(\mathcal{G}_0\) has a square-integrable envelope; that is, \(\E(\sup_{g\in\mathcal{G}_0}|g|)^2<\infty\).
  Then the estimator \(\hat\Omega=P_n\hat\phi^2\) is consistent for \(\Omega_{0}=P\phi_0^2\).
\end{corollary}

\begin{proof}
  We have the following expansion:
  \begin{align*}
    \hat\Omega &= P_n\bigg\{\bigg(\hat\ell-\frac{1-S}{\hat\alpha}\hat\theta\bigg)-\bigg(\ell_0-\frac{1-S}{\alpha_0}\theta_0\bigg)\bigg\}^2+2P_n\hat\phi\phi_0 + P_n\phi_0^2.
  \end{align*}
  The third term converges to \(\Omega_0\) in probability by the law of large numbers.
  The second term is bounded by two times the product of the square root of the first term and \(\Omega_0^{1/2}\) by the Cauchy-Schwarz inequality.
  Therefore, it remains to show that the first term is \(o_{P_0}(1)\).
  
  We have
  \begin{multline*}
    P_n\bigg\{\bigg(\hat\ell-\frac{1-S}{\hat\alpha}\hat\theta\bigg)-\bigg(\ell_0-\frac{1-S}{\alpha_0}\theta_0\bigg)\bigg\}^2\leq 2P_n(\hat\ell-\ell_0)^2 + 2\hat\alpha\bigg(\frac{\hat\theta}{\hat\alpha}-\frac{\theta_0}{\alpha_0}\bigg)^2\\
    \leq 2\sup_{g\in\mathcal{G}_0}|(P_n-P_0)g^2| + 2\|\hat\ell-\ell_0\|_{P_0}^2+ 2\hat\alpha\bigg(\frac{\hat\theta}{\hat\alpha}-\frac{\theta_0}{\alpha_0}\bigg)^2.
  \end{multline*}
  The third term in the rightmost part of the inequality above converges in probability to zero by the continuous mapping theorem and the consistency of \(\hat\theta\) and \(\hat\alpha\).
  The second term is \(o_{P_0}(1)\) as shown in the proof of Theorem~\ref{thm:asymptotics}.
  Finally, by Theorem~2.10.5 in \citetsuppmat{vandervaart2023weak}, the existence of a square-integrable envelope guarantees that the function class \(\{g^2:g\in\mathcal{G}_0\}\) is \(P_0\)-Glivenko-Cantelli.
  Therefore, the first term is \(o_{P_0}(1)\) by the uniform law of large numbers.
\end{proof}

\section{Asymptotic theory for estimators of bridge functions}
\label{sec:asymp-bridge-app}

In this section, we present useful results on the asymptotics of the proximal indirect comparison estimators in the simulation study.
In particular, we will derive their asymptotic variances under the assumption of parametric bridge functions.
Without loss of generality, we assume the parametric components have the same dimension.
Consider compact sets \(\mathrm{E},\Xi\subset\mathbb{R}^{k}\).

\begin{assumption}[Parametric bridge functions]
  \label{asn:parametric}
  For every \(P\in\P\), \(\H=\{h_{\eta_{0}}\}\) and \(\Q=\{q_{\xi_{0}}\}\) are singletons, where
  \(\eta_{0}\in\mathrm{E}\) and \(\xi_{0}\in\Xi\).
\end{assumption}

We choose some basis expansions \(b_{*}(z,x)\) and \(c_{*}(w,x)\) in \(\mathbb{R}^{k}\) and define the functions
\begin{align*}
  \psi^{h}_{\eta}(o) &= sb_{*}(z,x)\{\tilde{y}-h_{\eta}(w,x)\}, \\
  \psi^{q}_{\xi}(o) &= c_{*}(w,x)\{(1-s)-sq_{\xi}(z,x)\}.
\end{align*}
For a real-valued vector \(v\in\mathbb{R}^k\), let \(\|v\|\) denote its Euclidean norm.
The Z-estimators \(\hat{\eta}\in\mathrm{E}\) and \(\hat{\xi}\in\Xi\) are such that \(\|P_{n}\psi^{h}_{\hat{\eta}}\|=o_{P}(n^{-1/2})\) and \(\|P_{n}\psi^{q}_{\hat{\xi}}\|=o_{P}(n^{-1/2})\).
Note that the estimation of \(\xi_{0}\) need not involve nuisance models for the participation odds \(\Pr(S=0\mid W,X)/\Pr(S=1\mid W,X)\).
A similar point is raised in the estimation of parametric treatment bridge functions to account for unmeasured confounding \citetsuppmat{cui2024semiparametric}.

\begin{assumption}[Regularity conditions]
  \label{asn:if-simulation}
  \hfill
  \begin{enumerate}[label=(\roman*)]
  \item The map \(\eta\mapsto h_{\eta}(w,x)\) is continuous in \(\mathrm{E}\) for all \((w,x)\).
    \(\max_{1\leq j\leq k}\E\sup_{\eta\in\mathrm{E}}|\psi^h_{\eta,j}|<\infty\).
    The map \(\eta\mapsto \|\E\psi^{h}_{\eta}\|\) has a unique zero point \(\eta_{0}\in\mathrm{E}\).
    The map \(\eta\mapsto h_{\eta}(w,x)\) is differentiable at \(\eta_0\) for all \((w,x)\) with derivative \(\dot{h}_{\eta_0}\).
    The matrix \(E\dot\psi^h_{\eta_0}\) is invertible, where \(\dot\psi^h_{\eta_0}=-sb_{*}\dot{h}_{\eta_0}\).
    In some neighbourhood of \(\eta_0\) within \(\mathrm{E}\), the map \(\eta\mapsto\psi^h_\eta\) is \(M_h(o)\)-Lipschitz with \(M_h\in L_2(P)\).
  \item
    The map \(\xi\mapsto q_{\xi}(z,x)\) is continuous in \(\Xi\) for all \((z,x)\).
    \(\max_{1\leq j\leq k}\E\sup_{\xi\in\Xi}|\psi^q_{\xi,j}|<\infty\).
    The map \(\xi\mapsto \|\E\psi^{q}_{\xi}\|\) has a unique zero point \(\xi_{0}\in\Xi\).
    The map \(\xi\mapsto q_{\xi}(z,x)\) is differentiable at \(\xi_0\) for all \((z,x)\) with derivative \(\dot{q}_{\xi_0}\).
    The matrix \(E\dot\psi^q_{\xi_0}\) is invertible, where \(\dot\psi^q_{\xi_0}=-sc_{*}\dot{q}_{\xi_0}\).
    In some neighbourhood of \(\xi_0\) within \(\Xi\), the map \(\xi\mapsto\psi^q_\xi\) is \(M_q(o)\)-Lipschitz with \(M_q\in L_2(P)\).
  \end{enumerate}
\end{assumption}

\begin{proposition}
  \label{ppn:if-simulation}
  Suppose Assumptions~\ref{asn:parametric} and \ref{asn:if-simulation} hold. Then:
  \begin{enumerate}
  \item  The estimator \(\hat{\eta}\) is asymptotically normal with influence function \(-(\E\dot{\psi}^{h}_{\eta_{0}})^{-1}\psi^{h}_{\eta_{0}}\).
    The estimator \(\hat{\theta}_{h}\) is asymptotically normal with influence function
    \[
      \zeta^h_{\eta_0}(o)=\frac{1-s}{\alpha}\{h_{\eta_{0}}-\theta\} - \frac{1}{\alpha}\E(S\dot{h}_{\eta_{0}}^{\T})(\E\dot{\psi}^{h}_{\eta_{0}})^{-1}\psi^{h}_{\eta_{0}}.
    \]
  \item The estimator \(\hat{\xi}\) is asymptotically normal with influence function \(-(\E\dot{\psi}^{q}_{\xi_{0}})^{-1}\psi^{q}_{\xi_{0}}\).
    The estimator \(\hat{\theta}_{q}\) is asymptotically normal with influence function
    \[
      \zeta^q_{\xi_0}(o)=\frac{1}{\alpha}\big\{sq_{\xi_{0}}\tilde{y}-(1-s)\theta\big\} - \frac{1}{\alpha}\E(S\tilde{Y}\dot{q}_{\xi_{0}}^{\T})(\E\dot{\psi}^{q}_{\xi_{0}})^{-1}\psi^{q}_{\xi_{0}}.
    \]
  \end{enumerate}
\end{proposition}

\begin{proof}
  We show asymptotic properties of the estimator \(\hat\eta\) of \(\eta_0\).
  The continuity of \(h_\eta\), \(\max_{1\leq j\leq k}\E\sup_{\eta\in\mathrm{E}}|\psi^h_{\eta,j}|<\infty\), and the compactness of \(\mathrm{E}\) implies \(\sup_{\eta\in\mathrm{E}}\|(P_n-P)\psi^h_{\eta}\|=o_{P}(1)\) by Theorem~2.10.5 in \citetsuppmat{vandervaart2023weak}.
  Then \(\|P_n\psi^h_{\hat\eta}\|=o_{P}(1)\) and uniqueness of the zero point implies \(\|\hat\eta-\eta_0\|=o_{P}(1)\).
  This is consistency.
  The consistency of \(\hat\eta\), the Lipschitz condition on \(\psi^h_\eta\), the compactness of \(\mathrm{E}\), the differentiability of \(h_\eta\), and the invertibility of \(E\dot\psi^h_{\eta_0}\) establish the asymptotic linearity of \(\hat\eta\) by Theorems~2.5.6, 2.7.17, and 3.3.1 and Lemma~3.3.5 in \citetsuppmat{vandervaart2023weak}.
  The asymptotic linearity of \(\hat{\theta}_{h}\) can be seen by a first-order Taylor expansion around \(\eta_0\).

  The results for \(\hat\xi\) and \(\hat{\theta}_{q}\) can be obtained analogously under the corresponding conditions.
  Establishing the asymptotic linearity of \(\hat\theta_q\) requires an additional expansion around \(\alpha\).
\end{proof}

With mild regularity conditions on the influence functions \(\zeta^h_\eta\) and \(\zeta^q_\xi\), we obtain consistent estimators of the asymptotic variances of \(\hat\theta_h\) and \(\hat\theta_q\).

\begin{corollary}
  Suppose Assumptions~\ref{asn:parametric} and \ref{asn:if-simulation} hold.
  Furthermore, suppose the map \(\eta\mapsto \zeta^h_\eta(o)\) is continuous in \(\mathrm{E}\) for all \(o\) and that the map \(\xi\mapsto \zeta^q_\xi(o)\) is continuous in \(\Xi\) for all \(o\).
  Additionally, \(\max_{1\leq j\leq k}\E\sup_{\eta\in\mathrm{E}}(\zeta^h_{\eta,j})^2<\infty\) and \(\max_{1\leq j\leq k}\E\sup_{\xi\in\Xi}(\zeta^q_{\xi,j})^2<\infty\).
  Then:
  \begin{enumerate}
  \item \(\hat\Omega_{h} = P_n(\zeta^h_{\hat\eta})^2\) is a consistent estimator of the asymptotic variance \(\Omega_h=P(\zeta^h_{\eta_0})^2\) of \(\hat\theta_h\).
  \item \(\hat\Omega_{q} = P_n(\zeta^q_{\hat\xi})^2\) is a consistent estimator of the asymptotic variance \(\Omega_q=P(\zeta^q_{\xi_0})^2\) of \(\hat\theta_q\).
  \end{enumerate}
\end{corollary}

The estimators \(n^{-1/2}\hat\Omega_h^{1/2}\) and \(n^{-1/2}\hat\Omega_q^{1/2}\) were used to calculate standard errors of the corresponding proximal estimators in the simulation study.

\begin{figure}
  \centering
  \includegraphics[scale=1]{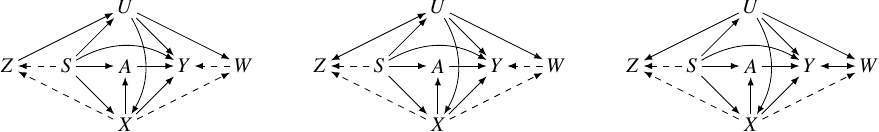}
  \caption{Acyclic directed mixed graphs compatible with Assumption~\ref{asn:identify}.}
  \label{fig:dag-add}
\end{figure}

\section{Existence and uniqueness of bridge functions}
\label{sec:bridge-app}

To make statistical inference on the target parameter, we rely on existence and uniqueness of the bridge functions \(h_{0}\) and \(q_{0}\).
For the sake of completness, we state sufficient conditions for this assumption.

We make use of the following completeness conditions on the observed data distribution, under which \(\H\) and \(\Q\) must be either empty sets or singletons.

\begin{assumption}[Completeness]
  \label{asn:completeness}
  \(P_{1}\)-almost surely:
  \begin{enumerate}[label=(\roman*)]
  \item\label{eqn:completeness-1}
    \(\E\{g(W,X)\mid Z,X,S=1\} = 0\) implies \(g(W,X) = 0\);
  \item\label{eqn:completeness-2}
    \(\E\{g(Z,X)\mid W,X,S=1\} = 0\) implies \(g(Z,X) = 0\).
  \end{enumerate}
\end{assumption}

Similar completeness assumptions appear in many works on proximal causal inference \citepsuppmat{cui2024semiparametric,tchetgentchetgen2024introduction}.
They use completeness assumptions to translate observed data parameters into causal parameters.
In our setup, this corresponds to viewing the identification formulas in Proposition~\ref{ppn:identification} directly as parameters of interest.
Then, with Assumption~\ref{asn:identify} and completeness assumptions on the densities \(p(u\mid W=w,X=x,S=1)\) and \(p(u\mid Z=z,X=x,S=1)\), we arrive at \(\H=\H^{\u}\) and \(\Q=\Q^{\u}\).
That is, under these alternative assumptions, the two parameters of interest will have the right causal interpretation.

We introduce some notations.
Let \(P_{1}\vert_{X=x}\) denote the conditional probability measure \(P_{1}(\cdot\mid X=x)\).
For every fixed \(x\), define the linear transformation \(T_{x}:L_{2}(W;P_{1}\vert_{X=x})\to L_{2}(Z;P_{1}\vert_{X=x})\) such that \((T_{x}h)(z)=\E\{h(W)\mid Z=z,X=x\}\).
The adjoint of \(T_{x}\) is then \(T_{x}^{*}:L_{2}(Z;P_{1}\vert_{X=x})\to L_{2}(W;P_{1}\vert_{X=x})\) such that \((T_{x}^{*}q)(w)=\E\{q(Z)\mid W=w,X=x\}\).
When \(T_{x}\) and \(T_{x}^{*}\) are compact operators, there exist orthonormal sequences \((f_{k})\) in \(L_{2}(Z;P_{1}\vert_{X=x})\) and \((g_{k})\) in \(L_{2}(W;P_{1}\vert_{X=x})\) and a positive sequence of real numbers \((\sigma_{k})\) such that \(T_xg_{k}=\sigma_{k}f_{k}\) and \(T_x^{*}f_{k}=\sigma_{k}g_{k}\) for all positive integers \(k\) (Theorem 15.16 in \citesuppmat{kress2014linear}).

\begin{assumption}[Regularity conditions]
  \label{asn:regularity}
  For all \(x\):
  \begin{enumerate}[label=(\roman*)]
  \item\label{eqn:regularity-1} \(T_{x}\) and \(T_{x}^{*}\) are compact operators;
  \item\label{eqn:regularity-2} \(\E(\tilde{Y}\mid Z,X=x,S=1)\in L_{2}(Z;P_{1}\vert_{X=x})\);
  \item\label{eqn:regularity-3} \(\{\Pr(S=1\mid W,X=x)\}^{-1}\Pr(S=0\mid W,X=x)\in L_{2}(W;P_{1}\vert_{X=x})\);
  \item\label{eqn:regularity-4} \(\sum_{k=1}^{\infty}\sigma_{k}^{-2}\big\vert\langle \E(\tilde{Y}\mid Z,X=x,S=1),f_{k}\rangle_{L_{2}(Z;P_{1}\vert_{X=x})}\big\vert^{2}<\infty\);
  \item\label{eqn:regularity-5} \(\sum_{k=1}^{\infty}\sigma_{k}^{-2}\big\vert\big\langle\{\Pr(S=1\mid W,X=x)\}^{-1}\Pr(S=0\mid W,X=x),g_{k}\rangle_{L_{2}(W;P_{1}\vert_{X=x})}\big\vert^{2}<\infty\).
  \end{enumerate}
\end{assumption}

\begin{proposition}[Identifiability of bridge functions]
  \label{ppn:solution}
  \hfill
  \begin{enumerate}
  \item Under Assumptions~\ref{asn:completeness}\ref{eqn:completeness-1}, \ref{asn:regularity}\ref{eqn:regularity-1}, \ref{asn:regularity}\ref{eqn:regularity-2}, and \ref{asn:regularity}\ref{eqn:regularity-4}, \(\H\) is nonempty and a singleton.
  \item Under Assumptions~\ref{asn:completeness}\ref{eqn:completeness-2}, \ref{asn:regularity}\ref{eqn:regularity-1}, \ref{asn:regularity}\ref{eqn:regularity-3}, and \ref{asn:regularity}\ref{eqn:regularity-5}, \(\Q\) is nonempty and a singleton.
  \end{enumerate}
\end{proposition}

\begin{proof}
  We only show the proof of the first index of Proposition~\ref{ppn:solution}.
  The singular value decomposition of \(T_{x}\) exists by Assumption~\ref{asn:regularity}\ref{eqn:regularity-1}.
  By definition, the nonemptiness of \(\H\) follows from the existence of a solution to the linear integral equation \((T_{x}h)(z)=\E(\tilde{Y}\mid Z=z,X=x,S=1)\) in the Hilbert space \(L_{2}(Z;P_{1}\vert_{X=x})\) for every \(x\).
  Applying Picard's Theorem (Theorem 15.18 in \citesuppmat{kress2014linear}), the equation for \(h\) has a solution due to Assumption~\ref{asn:regularity}\ref{eqn:regularity-2} and \ref{asn:regularity}\ref{eqn:regularity-4}.
  We will use proof by contradiction to show the second part of the statement.
  Suppose the contrary that there exist two solutions \(h\neq h^{\prime}\) to the equation in \(\H\), such that \(\E(h-h^{\prime}\mid Z,X=x,S=1)=0\) holds \(P_{1}(Z\mid X=x)\)-almost surely.
  Then by Assumption\ref{asn:completeness}\ref{eqn:completeness-1}, we must have \(h=h^{\prime}\) almost surely.
  The proof for the second index can be similarly obtained using the assumptions shown in Proposition~\ref{ppn:solution}.
\end{proof}

In general, the operators \(T_{x}\) and \(T_{x}^{*}\) are not compact operators.
A sufficient condition for Assumption~\ref{asn:regularity}\ref{eqn:regularity-1} is \(\iint p(w\mid z,X,S=1)p(z\mid w,X,S=1)\d w\d z < \infty\), \(P_{1}(X)\)-almost surely [see Example 2.3 in \citesuppmat{carrasco2007chapter}].
In this case, \(T_{x}\) is a Hilbert-Schmidt operator, which is guaranteed to be compact.


\section{Details of the simulated data example}
\label{sec:simulation-app}

\subsection{Underlying bridge functions}

The baseline covariates \(X\), the unobserved variables \(U\), the negative control outcomes \(W\), and the negative control treatments \(Z\) are multivariate.
The rest of the variables are univariate.
We use \(b\), \(\beta\), and \(B\) for scalar, vector, and matrix coefficients.
Their dimensions should be clear from the context.
We generated the data sequentially according to (excluding \(X\) and \(U\), which can follow arbitrary joint distributions):
\begin{align*}
  S\mid (X,U) &\sim \mathrm{Bernoulli}\{\mathrm{expit}(b_{s}+\beta_{sx}^{\T}X+\beta_{su}^{\T}U)\}, \\
  A\mid (X,S=1) &\sim \mathrm{Bernoulli}(b_{a}), 0<b_{a}<1,\\
  Z\mid (X,U,S=1) &\sim \mathrm{Normal}(\beta_{z}+B_{zu}U+B_{zx}X,\Sigma_{z}), \\
  W\mid (X,U) &\sim \mathrm{Normal}(\beta_{w}+B_{wu}U+B_{wx}X,\Sigma_{w}), \\
  Y\mid (W,A,X,U,S=1) &\sim \mathrm{Normal}(b_{y}+b_{ya}A + \beta_{yu}^{\T}U + \beta_{y(au)}^{\T}(AU) + \beta_{yx}^{\T}X\\
  &\hphantom{\sim \mathrm{Normal}(}+ \beta_{yw}^{\T}W+\beta_{y(aw)}^{\T}(AW),\sigma_{y}^{2}).
\end{align*}

The conditional distribution \(W\mid (A,X,U,S=1)\) is the same as the conditional distribution of \(W\mid (X,U,S=1)\).
The conditional expectation
\begin{align*}
  \MoveEqLeft \E(Y\mid A=1,X,U,S=1)-\E(Y\mid A=0,X,U,S=1) \\
  &= \E\{\E(Y\mid W,A=1,X,U,S=1)-\E(Y\mid W,A=0,X,U,S=1)\mid X,U,S=1\} \\
  &= b_{ya}+\beta_{y(au)}^{\T}U+\beta_{y(aw)}^{\T}\E(W\mid X,U,S=1) \\
  &= b_{ya}+\beta_{y(aw)}^{\T}\beta_{w} + (\beta_{y(au)}^{\T}+\beta_{y(aw)}^{\T}B_{wu})U + \beta_{y(aw)}^{\T}B_{wx}X.
\end{align*}
Let the outcome bridge function \(h(W,X) = \eta_{0} + \eta_{w}^{\T}W + \eta_{x}^{\T}X\), then
\begin{align*}
  \E\{h(W,X)\mid X,U,S=1\} &= \eta_{0} + \eta_{x}^{\T}X + \eta_{w}^{\T}\E(W\mid X,U,S=1) \\
                        &= \eta_{0} + \eta_{w}^{\T}\beta_{w} + (\eta_{x}^{\T}+\eta_{w}^{\T}B_{wx})X + \eta_{w}^{\T}B_{wu}U.
\end{align*}
Comparing the coefficients in the two expressions, we have the following system of equations:
\begin{align*}
  \eta_{0} + \eta_{w}^{\T}\beta_{w} &= b_{ya}+\beta_{y(aw)}^{\T}\beta_{w} \\
  \eta_{x}^{\T}+\eta_{w}^{\T}B_{wx} &= \beta_{y(aw)}^{\T}B_{wx} \\
  \eta_{w}^{\T}B_{wu} &= \beta_{y(au)}^{\T}+\beta_{y(aw)}^{\T}B_{wu},
\end{align*}
so the parameters of the bridge function are
\begin{align*}
  \eta_{0} &= b_{ya} - \beta_{w}^{\T}B_{wu}^{-\T}\beta_{y(au)}, \\
  \eta_{x} &= - B_{wx}^{\T}B_{wu}^{-\T}\beta_{y(au)},\\
  \eta_{w} &= B_{wu}^{-\T}\beta_{y(au)} + \beta_{y(aw)}.
\end{align*}

The probability ratio
\[
  \frac{\Pr(S=0\mid X,U)}{\Pr(S=1\mid X,U)} = \exp(-b_{s}-\beta_{sx}^{\T}X-\beta_{su}^{\T}U).
\]
Let the participation bridge function be
\[
  q(Z,X) = \exp(\xi_{0} + \xi_{z}^{\T}Z + \xi_{x}^{\T}X), 
\]
then
\begin{align*}
  \E\{q(Z,X)\mid X,U,S=1\} &= \exp(\xi_{0} + \xi_{x}^{\T}X)\E\{\exp(\xi_{z}^{\T}Z)\mid X,U,S=1\} \\
                        &= \exp(\xi_{0} + \xi_{x}^{\T}X)\exp\bigg\{\xi_{z}^{\T}(\beta_{z}+B_{zu}U+B_{zx}X)+\frac{1}{2}\xi_{z}^{\T}\Sigma_{z}\xi_{z}\bigg\} \\
                        &= \exp\bigg\{\xi_{0}+\xi_{z}^{\T}\beta_{z}+\frac{1}{2}\xi_{z}^{\T}\Sigma_{z}\xi_{z}+(\xi_{x}^{\T}+\xi_{z}^{\T}B_{zx})X+\xi_{z}^{\T}B_{zu}U\bigg\},
\end{align*}
where we have used the moment generating function of the conditional distribution \(Z\mid (X,U,S=1)\).
Comparing the coefficients in the two expressions, we have the following system of equations:
\begin{align*}
  \xi_{0}+\xi_{z}^{\T}\beta_{z}+\frac{1}{2}\xi_{z}^{\T}\Sigma_{z}\xi_{z} &= -b_{s} \\
  \xi_{x}^{\T}+\xi_{z}^{\T}B_{zx} &= -\beta_{sx}^{\T} \\
  \xi_{z}^{\T}B_{zu} &= -\beta_{su}^{\T},
\end{align*}
so the parameters of the bridge function are
\begin{align*}
  \xi_{0} &= - b_{s} +\beta_{z}^{\T}B_{zu}^{-\T}\beta_{su}-\frac{1}{2}\beta_{su}^{\T}B_{zu}^{-1}\Sigma_{z}B_{zu}^{-\T}\beta_{su}, \\
  \xi_{x} &= -\beta_{sx} + B_{zx}^{\T}B_{zu}^{-\T}\beta_{su},\\
  \xi_{z} &= -B_{zu}^{-\T}\beta_{su}.
\end{align*}

\subsection{Justification for the cubic of basis}

Suppose we obtain an estimator of \(\xi_{0}\) without using the cubic of the basis \(c(w,x)\), so that
\[
  \tilde{\xi} = \arg\underset{\xi}{\min}\;\bigg\|\frac{1}{n}\sum_{i=1}^{n}c(W_{i},X_{i})\{S_{i}q_{\xi}(Z_{i},X_{i})-(1-S_{i})\}\bigg\|^{2}.
\]
The first-order condition of the optimization problem gives
\[
  2\bigg\{\frac{1}{n}\sum_{i=1}^{n}S_{i}q_{\tilde{\xi}}(Z_{i},X_{i})c(W_{i},X_{i})c^{\T}(W_{i},X_{i})\bigg\}\bigg\{\frac{1}{n}\sum_{i=1}^{n}c(W_{i},X_{i})\{S_{i}q_{\tilde{\xi}}(Z_{i},X_{i})-(1-S_{i})\}\bigg\}=0.
\]
Since the matrix in the first pair of braces is almost surely nonsingular, we have that with probability 1,
\begin{equation}
  \label{eqn:first-order}
  \frac{1}{n}\sum_{i=1}^{n}c(W_{i},X_{i})\{S_{i}q_{\tilde{\xi}}(Z_{i},X_{i})-(1-S_{i})\}=0.
\end{equation}
We compare the two estimators
\begin{align*}
  \tilde{\theta} &= \frac{1}{n}\sum_{i=1}^{n}\bigg[\frac{S_{i}}{\hat{\alpha}}q_{\tilde{\xi}}(Z_{i},X_{i})\{\tilde{Y}_{i}-h_{\hat{\eta}}(W_{i},X_{i})\}+\frac{1-S_{i}}{\hat{\alpha}}h_{\hat{\eta}}(W_{i},X_{i})\bigg], \\
\tilde{\theta}_{q} &=\frac{1}{n_{0}}\sum_{i:S_{i}=1}q_{\tilde{\xi}}(Z_{i},X_{i})\tilde{Y}_{i},
\end{align*}
which are the same as \(\hat{\theta}\) and \(\hat{\theta}_{q}\), except \(\hat{\xi}\) is replaced by \(\tilde{\xi}\).
Their difference is
\[
  \tilde{\theta}_{q}-\tilde{\theta} = \frac{1}{n}\sum_{i=1}^{n}\frac{1}{\hat{\alpha}}\{S_{i}q_{\tilde{\xi}}(Z_{i},X_{i})-(1-S_{i})\}h_{\hat{\eta}}(W_{i},X_{i}).
\]
Because \(h_{\hat{\eta}}(w,x)=\hat{\eta}^{\T}c(w,x)\) is a linear combination of \(c(w,x)\), the observation in \eqref{eqn:first-order} shows that \(\tilde{\theta}_{q}=\tilde{\theta}\) almost surely.
However, this can be circumvented by using a nonlinear transformation of \(c(w,x)\) as the basis function to estimate \(\xi_{0}\), as was done in the simulation study, where we raised \(c(w,x)\) to the third power elementwise.

\subsection{Additional experiments under assumption violations}

In experiment 5, we investigated the behaviour of proximal indirect comparison estimators in the absence of unmeasured effect modifiers by setting \(U\) to a zero vector.
To understand the impact of violations of the proxy assumptions (Assumption~\ref{asn:identify}\ref{eqn:identify-zy} and \ref{asn:identify}\ref{eqn:identify-ws} in the main text), we replaced the conditional distribution of \(Y\) with \(Y\mid (Z,W,A,U,S=0)\sim \mathrm{Normal}(0.5-A+U^{\T}1+AU^{\T}1+X^{\T}1+W^{\T}1+AW^{\T}1+Z^{\T}1+AZ^{\T}1,0.5^{2})\) in experiment 7 and the condition distribution of \(W\) with \(W\mid (X,S)\sim \mathrm{Normal}(S1+X+U,0.25\mathrm{Id})\) in experiment 8.
In experiment 9, we simulated \(U\sim\mathrm{Uniform}([-1,0])\) as a scalar-valued random variable but maintained the proxies as vectors, so that the bridge functions are no longer uniquely identified.
The importance of the existence of the bridge functions was studied in experiment 11 by simulating \(Z\) from \(Z\mid (U,X,S=0)\sim\mathrm{Normal}(0.05U+X,0.25\mathrm{Id})\), making \(Z\) nearly uncorrelated with \(U\) given \(X\) in the source RCT.
Likewise in experiment 12, we simulated \(W\) from \(W\mid (U,X)\sim \mathrm{Normal}(0.05U+X,0.25\mathrm{Id})\) so that \(W\) is nearly uncorrelated with \(U\) given \(X\).
In experiment 13, we examined the effect on near violation of positivity by changing the conditional probability \(\Pr(S=1\mid U,X)\) to \(\mathrm{expit}(-0.675+0.5X^{\T}1+2.5U^{\T}1)\) so that the participation odds \(\Pr(S=0\mid U,X)/\Pr(S=1\mid U,X)\) is large, as the coefficient of \(U\) is dispropotionally large.
Finally in experiments 6 and 10, the data were simulated as in experiments 5 and 9, whereas the bridge functions were estimated with ridge regularization.
That is, the fitted bridge functions were \(h_{\hat{\eta}_{\lambda}}(W,X)\) and \(q_{\hat{\xi}_\lambda}(Z,X)\) where
\begin{align*}
  \hat{{\eta}}_{\lambda} &=\arg\underset{{\eta}'}{\min}\;\bigg\|\frac{1}{n_{1}}\sum_{i:S_{i}=1}b(Z_{i},X_{i})\{\tilde{Y}_{i}-{h}_{\eta'}(W_{i},X_{i})\}\bigg\|^{2}+\lambda_{h}(\eta')^{\T}D_{h}\eta', \\
  \hat{{\xi}}_{\lambda} &=\arg\underset{{\xi}'}{\min}\;\bigg\|\frac{1}{n}\sum_{i=1}^{n}\{c(W_{i},X_{i})\}^{3}\{S_{i}{q}_{\xi'}(Z_{i},X_{i})-(1-S_{i})\}\bigg\|^{2}+\lambda_{q}(\xi')^{\T}D_{q}\xi',
\end{align*}
with fixed regularization parameters \(\lambda_{h}=\lambda_{q}=10^{-4}\) and \(D_{h}\) and \(D_{q}\) being identity matrices of appropriate dimensions with their upper left corners changed to zero, so that the intercept is unpenalized.
All results from the additional simulation studies are displayed in Tables~\ref{tab:sim-add} and \ref{tab:sim-add-2}.

\begin{table}
  \caption{Additional simulation results of experiments 5--13 with sample size \(n=1000\).}
  \label{tab:sim-add}
  \footnotesize\centering
  {
    \begin{tabular}[t]{lllS[]S[]rrr}
      \toprule
      {\(n\)} & {Experiment} & {Estimator} & {Mean} & {Bias} & {RMSE} & {SE} & {Coverage}\\
      \midrule
      \(1000\) & 5 & \(\hat{\theta}_{h}\) & 0.40 & -3.29 & 60.24 & 585.37 & 100.00\\
              &  & \(\hat{\theta}_{q}\) & 0.38 & -25.55 & 9.79 & 1318.56 & 97.39\\
              &  & \(\hat{\theta}\) & 0.17 & -232.71 & 140.78 & 65.29 & 98.59\\
      \cmidrule{2-8}
              & 6 & \(\hat{\theta}_{h}\) & 0.40 & -8.56 & 3.39 & 63.76 & 99.80\\
              &  & \(\hat{\theta}_{q}\) & 0.39 & -12.37 & 3.36 & 42.13 & 99.50\\
              &  & \(\hat{\theta}\) & 0.39 & -13.32 & 3.36 & 3.31 & 94.70\\
      \cmidrule{2-8}
              & 7 & \(\hat{\theta}_{h}\) & -2.84 & -93.51 & 4.87 & 4.82 & 95.30\\
              &  & \(\hat{\theta}_{q}\) & -2.87 & -116.60 & 6.41 & 6.27 & 95.70\\
              &  & \(\hat{\theta}\) & -2.84 & -93.01 & 5.83 & 5.78 & 94.80\\
      \cmidrule{2-8}
              & 8 & \(\hat{\theta}_{h}\) & -5.45 & -2802.40 & 52.29 & 44.69 & 90.60\\
              &  & \(\hat{\theta}_{q}\) & -2.01 & 631.36 & 33.84 & 1.65 & 5.48\\
              &  & \(\hat{\theta}\) & -5.13 & -2487.30 & 53.51 & 28.60 & 55.01\\
      \cmidrule{2-8}
              & 9 & \(\hat{\theta}_{h}\) & -1.83 & -169.06 & 43.99 & 379.44 & 100.00\\
              &  & \(\hat{\theta}_{q}\) & -1.70 & -37.91 & 5.68 & 150.45 & 96.70\\
              &  & \(\hat{\theta}\) & -1.86 & -206.71 & 56.19 & 28.78 & 97.49\\
      \cmidrule{2-8}
              & 10 & \(\hat{\theta}_{h}\) & -1.65 & 3.76 & 2.39 & 28.39 & 99.70\\
              &  & \(\hat{\theta}_{q}\) & -1.66 & -5.54 & 2.56 & 27.70 & 99.50\\
              &  & \(\hat{\theta}\) & -1.66 & -0.08 & 2.51 & 2.47 & 94.20\\
      \cmidrule{2-8}
              & 11 & \(\hat{\theta}_{h}\) & -2.73 & -79.91 & 72.12 & 725.79 & 100.00\\
              &  & \(\hat{\theta}_{q}\) & -2.52 & 121.77 & 10.02 & 24.72 & 90.67\\
              &  & \(\hat{\theta}\) & -2.64 & 8.94 & 73.71 & 41.25 & 95.49\\
      \cmidrule{2-8}
              & 12 & \(\hat{\theta}_{h}\) & -1.09 & 85.97 & 60.62 & 582.12 & 100.00\\
              &  & \(\hat{\theta}_{q}\) & -1.10 & 84.36 & 6.68 & 947.45 & 96.40\\
              &  & \(\hat{\theta}\) & -1.31 & -134.55 & 70.33 & 40.68 & 99.00\\
      \cmidrule{2-8}
              & 13 & \(\hat{\theta}_{h}\) & -2.67 & -111.00 & 31.82 & 44.00 & 94.10\\
              &  & \(\hat{\theta}_{q}\) & -2.22 & 336.45 & 19.65 & 7.44 & 44.10\\
              &  & \(\hat{\theta}\) & -2.43 & 130.62 & 29.32 & 18.92 & 83.47\\
      \bottomrule
    \end{tabular}}

\medskip
  {Bias: Monte-Carlo bias, \(10^{-3}\); RMSE: root mean squared error, \(10^{-1}\); SE: average of standard error estimates, \(10^{-1}\); Coverage: \(95\%\) confidence interval coverage, \(\%\).
  Experiment 5: no unmeasured effect modifiers; experiment 6: no unmeasured effect modifiers with ridge regularization; experiment 7: violation of \(Y\)-\(Z\) conditional independence, experiment 8: violation of \(W\)-\(S\) conditional independence; experiment 9: over-identification of bridge functions; experiment 10: over-identification of bridge functions with ridge regularization; experiment 11: near violation of \(Z\)-\(U\) relevance; experiment 12: near violation of \(W\)-\(U\) relevance; experiment 13: near violation of positivity of \(\Pr(S=1\mid U,X)\).}
\end{table}

\begin{table}
  \caption{Additional simulation results of experiments 5--13 with sample size \(n=2000\).}
  \label{tab:sim-add-2}
  \footnotesize\centering
  {\begin{tabular}[t]{lllS[]S[]rrr}
    \toprule
    {\(n\)} & {Experiment} & {Estimator} & {Mean} & {Bias} & {RMSE} & {SE} & {Coverage}\\
    \midrule
    \(2000\) & 5 & \(\hat{\theta}_{h}\) & 0.56 & 154.81 & 37.69 & 367.80 & 100.00\\
            &  & \(\hat{\theta}_{q}\) & 0.40 & -2.34 & 6.72 & 59.25 & 98.99\\
            &  & \(\hat{\theta}\) & 0.42 & 9.49 & 91.75 & 39.15 & 98.79\\
    \cmidrule{2-8}
            & 6 & \(\hat{\theta}_{h}\) & 0.42 & 11.12 & 2.45 & 84.90 & 99.80\\
            &  & \(\hat{\theta}_{q}\) & 0.41 & 6.37 & 2.42 & 33.26 & 99.80\\
            &  & \(\hat{\theta}\) & 0.41 & 4.56 & 2.42 & 2.32 & 92.80\\
    \cmidrule{2-8}
            & 7 & \(\hat{\theta}_{h}\) & -2.85 & -104.03 & 3.59 & 3.37 & 93.80\\
            &  & \(\hat{\theta}_{q}\) & -2.87 & -124.49 & 4.40 & 4.19 & 95.40\\
            &  & \(\hat{\theta}\) & -2.86 & -110.04 & 4.08 & 3.90 & 94.50\\
    \cmidrule{2-8}
            & 8 & \(\hat{\theta}_{h}\) & -5.60 & -2955.11 & 42.89 & 31.01 & 84.30\\
            &  & \(\hat{\theta}_{q}\) & -2.35 & 293.72 & 39.12 & 2.15 & 4.41\\
            &  & \(\hat{\theta}\) & -5.29 & -2649.47 & 51.14 & 32.02 & 62.07\\
    \cmidrule{2-8}
            & 9 & \(\hat{\theta}_{h}\) & -1.58 & 77.82 & 20.55 & 144.31 & 99.70\\
            &  & \(\hat{\theta}_{q}\) & -1.68 & -24.05 & 4.03 & 548.32 & 97.00\\
            &  & \(\hat{\theta}\) & -1.67 & -16.70 & 26.86 & 17.81 & 98.30\\
    \cmidrule{2-8}
            & 10 & \(\hat{\theta}_{h}\) & -1.66 & -3.03 & 1.64 & 18.46 & 99.20\\
            &  & \(\hat{\theta}_{q}\) & -1.67 & -8.20 & 1.78 & 24.66 & 99.40\\
            &  & \(\hat{\theta}\) & -1.66 & -7.20 & 1.73 & 1.74 & 95.20\\
    \cmidrule{2-8}
            & 11 & \(\hat{\theta}_{h}\) & -2.78 & -131.67 & 75.44 & 528.18 & 100.00\\
            &  & \(\hat{\theta}_{q}\) & -2.51 & 136.13 & 8.89 & 1529.49 & 92.30\\
            &  & \(\hat{\theta}\) & -2.59 & 51.16 & 67.84 & 27.79 & 92.29\\
    \cmidrule{2-8}
            & 12 & \(\hat{\theta}_{h}\) & -1.08 & 98.20 & 43.26 & 293.16 & 100.00\\
            &  & \(\hat{\theta}_{q}\) & -1.11 & 71.73 & 4.91 & 494.91 & 97.29\\
            &  & \(\hat{\theta}\) & -0.93 & 249.01 & 68.86 & 29.61 & 98.99\\
    \cmidrule{2-8}
            & 13 & \(\hat{\theta}_{h}\) & -2.58 & -17.43 & 12.05 & 11.62 & 94.80\\
            &  & \(\hat{\theta}_{q}\) & -2.61 & -53.22 & 17.31 & 11.60 & 51.69\\
            &  & \(\hat{\theta}\) & -2.66 & -99.47 & 17.48 & 15.17 & 88.39\\
    \bottomrule
  \end{tabular}}

\medskip
   {Bias: Monte-Carlo bias, \(10^{-3}\); RMSE: root mean squared error, \(10^{-1}\); SE: average of standard error estimates, \(10^{-1}\); Coverage: \(95\%\) confidence interval coverage, \(\%\).
   Experiment 5: no unmeasured effect modifiers; experiment 6: no unmeasured effect modifiers with ridge regularization; experiment 7: violation of \(Y\)-\(Z\) conditional independence, experiment 8: violation of \(W\)-\(S\) conditional independence; experiment 9: over-identification of bridge functions; experiment 10: over-identification of bridge functions with ridge regularization; experiment 11: near violation of \(Z\)-\(U\) relevance; experiment 12: near violation of \(W\)-\(U\) relevance; experiment 13: near violation of positivity of \(\Pr(S=1\mid U,X)\).}
\end{table}

\section{Details of the real data example}
\label{sec:analysis-app}

Let \(m(a,x,s)=\E(Y\mid A=a,X=x,S=s)\) and by an abuse of notation \(p(x)=\Pr(S=1\mid X=x)\), \(e(a\mid x,s)=\Pr(A=a\mid X=x,S=s)\).
We assumed linear models for \(m\) and \(\log\{p/(1-p)\}\).
Let \(\tilde{X}\) denote the design vector without intercept from the baseline adjusting variables \(X\) with numerical variables transformed into the orthogonal cubic basis and categorical variables transformed into dummy variables.
The bridge functions were assumed to follow the parametric forms \(h(w,x)=h_{\eta}(w,x)=\eta^{\T}\tilde{c}(w,x)\) and \(q(z,x)=q_{\xi}(z,x)=\xi^{\T}\tilde{b}(z,x)\), where \(\tilde{c}(w,x)=(1,w,\tilde{x}^{\T})^{\T}\) and \(\tilde{b}(z,x)=(1,z,\tilde{x}^{\T})^{\T}\).

The linear parameters in the bridge functions are fitted using the ridge-regularized generalized method of moment such that
\begin{align*}
  \hat{\eta}_{\lambda} &= \arg\min_{\eta}\bigg\|\frac{1}{n_{1}}\sum_{i:S_{i}=1}\tilde{b}(Z_{i},X_{i})\{\tilde{Y}_{i}-h_{\eta}(W_{i},X_{i})\}\bigg\|^{2} + \lambda_{h,n}\eta^{\T}D_{h}\eta, \\
  \hat{\xi}_{\lambda} &= \arg\min_{\xi}\bigg\|\frac{1}{n}\sum_{i=1}^{n}\tilde{c}(W_{i},X_{i})\{S_{i}q_{\xi}(Z_{i},X_{i})-(1-S_{i})\}\bigg\|^{2} + \lambda_{q,n}\xi^{\T}D_{q}\xi,
\end{align*}
where \(D_{h}\) and \(D_{q}\) are identity matrices of appropriate dimensions with their upper left corners changed to zero.
The data-adaptive regularization factors \(\lambda_{h,n}\) and \(\lambda_{q,n}\) are chosen with \(10\)-fold cross validation from a prespecified grid.
The models for \(m(a,\cdot,s)\) are fitted separately on the SCALE and STEP-2 samples as well as on each treatment arm to allow for full interaction between \(\{A,S\}\) and the other variables.
Then the modified target population ATE estimator from \citetsuppmat{dahabreh2020extending} for \(\theta\) is
\[
  \frac{1}{n}\sum_{i=1}^{n}\bigg[\frac{S_{i}}{\hat{\alpha}}\frac{1-\hat{p}(X_{i})}{\hat{p}(X_{i})}\frac{(2A_{i}-1)}{e(A_{i}\mid S_{i})}\{Y_{i}-\hat{m}(A_{i},X_{i},S_{i})\}+\frac{1-S_{i}}{\hat{\alpha}}\{\hat{m}(1,X_{i},S_{i})-\hat{m}(0,X_{i},S_{i})\}\bigg],
\]
and the modified standard doubly robust ATE estimator from \citetsuppmat{bang2005doubly} for \(\E\{Y(-1)-Y(0)\mid S=0\}\) is
\[
  \frac{1}{n_{0}}\sum_{i:S_{i}=0}\bigg[\frac{(-2A_{i}-1)}{e(A_{i}\mid 0)}\{Y_{i}-\hat{m}(A_{i},X_{i},0)\}+\{\hat{m}(-1,X_{i},0)-\hat{m}(0,X_{i},0)\}\bigg].
\]
For comparison, we also computed the ATE estimate for \(\E\{Y(1)-Y(0)\mid S=1\}\) by
\[
  \frac{1}{n_{1}}\sum_{i:S_{i}=1}\bigg[\frac{(2A_{i}-1)}{e(A_{i}\mid 1)}\{Y_{i}-\hat{m}(A_{i},X_{i},1)\}+\{\hat{m}(1,X_{i},1)-\hat{m}(0,X_{i},1)\}\bigg].
\]
The last two estimates are reported along with \(95\%\) confidence intervals in Table~\ref{tab:ic-app} for reference.

\begin{table}
  \caption{Additional estimates from the indirect comparison analysis with SCALE and STEP-2.}
  \label{tab:ic-app}
  \footnotesize\centering
   {\begin{tabular}{lcc}
   Estimand & Estimate & \(95\%\)-CI \\
   \(\E\{Y(-1)-Y(0)\mid S=0\}\) & \(-6.90\) & \((-7.68,-6.11)\) \\
   \(\E\{Y(1)-Y(0)\mid S=1\}\) & \(-3.97\) & \((-4.71,-3.22)\) \\
   \end{tabular}}

 \medskip
 {The estimands are direct comparisons of the treatments administered in the respective trials.}
\end{table}

\section{Handling missing outcome}
\label{sec:missing-app}

\subsection{Identifiability}
In our discussion thus far, we have ignored a common issue in many RCTs: study participants are typically followed over a period of time, during which some may drop out before the end of study, and their outcomes are not recorded.
When the dropout mechanism is not missing completely at random, applying the proximal indirect comparison estimator from previous sections to the nonmissing population may not identify the full-compliance ATE in the target RCT \(\theta\) due to potential selection bias.
In this section, we propose an estimator which correctly identifies the target parameter \(\theta\) under a missing-at-random (MAR) dropout pattern.

The binary missingness indicator \(\Delta\) takes the value \(0\) when a study participant's outcome information is missing.
Let the conditional probability of no dropout from the source trial be \(\pi(Z,W,A,X)=\Pr(\Delta=1\mid Z,W,A,X,S=1)\).
We assume that the missingness in the source trial is noninformative of the outcome, conditioning on all other observed variables, which is formalized below.

\begin{assumption}[Missing at random]
  \label{asn:mar}
  \hfill
  \begin{enumerate}[label=(\roman*)]
  \item\label{eqn:mar} \(\Delta\indep Y\mid (Z,W,A,X,S=1)\);
  \item\label{eqn:positivity-mar}\(\pi(Z,W,A,X)>0\) whenever \(\Pr(S=1\mid Z,W,A,X)>0\).
  \end{enumerate}
\end{assumption}

In particular, Assumption~\ref{asn:mar} requires that the unobserved effect modifiers \(U\) do not directly affect the missing pattern.
They are allowed to have an indirect effect through the proxies and the baseline covariates, upon controlling for which the missingness is ignorable.
If the outcome is MAR, one can devise augmented estimators from the influence functions of the target parameter \(\theta\) defined on the data without missingness \citepsuppmat{tsiatis2006semiparametric}.
If missing outcomes are also present in the target RCT, the identifiability of the ATE \(\theta\) comparing treatments \(A=0\) and \(A=-1\) relying only on randomization is lost.

The observed data model subject to missingness \(\P^{\c}\) is the collection of distributions over \(O^{\c}=(S,S\Delta,SA,X,W,SZ,S\Delta Y)\) such that \(P^{\c}(O^{\c}, S\Delta=1)=\pi(Z,W,A,X)P(O,S=1)\) and \(P^{\c}(O^{\c},S=0)=P(O,S=0)\) for all \(P\in\P\).
In this sense, we can write every \(P^{\c}\in\P^{\c}\) as a function \(P^{\c}(P)\).
The definition of the sets of bridge functions \(\H\) and \(\Q\) is valid for the model \(\P^{\c}\) under MAR, which immediately makes the target parameter identifiable.
Note that conditional mean of the outcome among subjects without missingness in the source trial \(\E(Y\mid \Delta=1,Z=z,W=w,A=a,X=x,S=1)\) is the same as the conditional mean \(\mu(z,w,a,x)=\E(Y\mid Z=z,W=w,A=a,X=x,S=1)\) where the outcome is always observed.

\subsection{Estimation}

\begin{proposition}
  \label{ppn:eif-mar}
  Suppose Assumption~\ref{asn:mar} holds.
  For \(P_0\in\P\) under Assumption~\ref{asn:tangent}, the efficient influence function of the target parameter \(\theta_0\) at \(P^{\c}_0\in\P^{\c}\) is
  \begin{multline*}
  \varphi^{\c}_{0}(o) = \frac{s}{\alpha_0}q_{0}(z,x)\frac{\delta(2a-1)}{\pi_0(z,w,a,x)e_0(a\mid x)}\{y-\mu_0(z,w,a,x)\} \\
              +\frac{s}{\alpha_0}q_{0}(z,x)\{\mu_0(z,w,1,x)-\mu_0(z,w,0,x)-h_{0}(w,x)\}+\frac{1-s}{\alpha_0}\{h_0(w,x)-\theta_0\}.
  \end{multline*}
\end{proposition}

\begin{proof}
  Recall that in the proof of Proposition~\ref{ppn:eif}, we have derived the tangent space \(\dot{\P}_0\) and it orthogonal complement \(\Lambda\).
  Therefore, the translation \(\varphi(o)+\Lambda\) is the space of all influence functions of the parameter \(\theta_0\) at \(P_0\) under the model \(\P\).
  More explicitly, the influence functions share the form
  \[
    \tilde{\phi}_{0}(o;c)=\varphi_0(o)+sc(z,w,x)\{a-e_0(1\mid x)\},
  \]
  where \(c\in L_2(P_0^1)\) is arbitrary.
  Following Example 25.43 in \citetsuppmat{vandervaart1998asymptotic}, all influence functions of \(\theta_0\) at \(P^{\c}\) under the model \(\P^{\c}\) can be characterized as
  \begin{align*}
    \tilde{\phi}_0^{\c}(o;c,b) &= \bigg\{\frac{s\delta}{\pi_0(z,w,a,x)}+(1-s)\bigg\}\tilde{\phi}_0(o;c)+sb(z,w,a,x)\{\delta-\pi_0(z,w,a,x)\} \\
                               &=\bigg\{\frac{s\delta}{\pi_0(z,w,a,x)}+(1-s)\bigg\}\big[\varphi_0(o)+sc(z,w,x)\{a-e_0(1\mid x)\}]\big]\\
                               &\hphantom{=}\quad +sb(z,w,a,x)\{\delta-\pi_0(z,w,a,x)\}\\
                               &= \frac{s}{\alpha_0}\frac{\delta}{\pi_0(z,w,a,x)}q_0(z,x)\frac{2a-1}{e_0(a\mid x)}\{y-\mu_0(z,w,a,x)\} \\
                               &\phantom{=}\quad + \frac{s}{\alpha_0}\frac{\delta}{\pi_0(z,w,a,x)}q_0(z,x)\{\mu_0(z,w,1,x)-\mu_0(z,w,0,x)-h_0(w,x)\} \\
                               &\phantom{=}\quad +\frac{1-s}{\alpha_0}\{h_0(w,x)-\theta_0\} + \frac{s\delta}{\pi_0(z,w,a,x)}c(z,w,x)\{a-e_0(1\mid x)\} \\
                               &\hphantom{=}\quad +sb(z,w,a,x)\{\delta-\pi_0(z,w,a,x)\},
  \end{align*}
  where \(b\in L_2(P_0^1)\) is arbitrary.
  To find the efficient influence function, we first optimize over \(b\).
  This is equivalent to calculating the projection of 
  \[
    \bigg\{\frac{s\delta}{\pi_0(z,w,a,x)}+(1-s)\bigg\}\tilde{\phi}_0(o;c)
  \]
  onto \((\Lambda^{\c})^\perp\), where
  \[
    \Lambda^{\c}=\{sb(z,w,a,x)\{\delta-\pi_0(z,w,a,x)\}:b(z,w,a,x)\in L_{2}(P^1_{0})\}.
  \]
  Suppose the projection has the form \(s\{\delta-\pi_0(z,w,a,x)\}b^{*}(z,w,a,x)\).
  Then the function \(b^{*}\) satisfies the equation
  \begin{multline*}
    \E_{P_0}\bigg(\bigg[\frac{\Delta \tilde{\phi}_0(O;c)}{\pi_0(Z,W,A,X)} - \{\Delta-\pi_0(Z,W,A,X)\}b^{*}(Z,W,A,X)\bigg]\\
    \{\Delta-\pi_0(Z,W,A,X)\}\biggm\vert Z,W,A,X,S=1\bigg)=0.
  \end{multline*}
  The solution is
  \begin{multline*}
    b^{*}(z,w,a,x) = \frac{q_0(z,x)}{\alpha_0\pi_0(z,w,a,x)}\{\mu_0(z,w,1,x)-\mu_0(z,w,0,x)-h_0(w,x)\}\\
    + \frac{c(z,w,x)}{\pi_0(z,w,a,x)}\{a-e_0(1\mid x)\},
  \end{multline*}
  which gives the projection
  \begin{align}
    \tilde{\phi}_0^{\c}(o;c,-b^*) &= \frac{s}{\alpha_0}q_{0}(z,x)\frac{2a-1}{e_0(a\mid x)}\frac{\delta}{\pi_0(z,w,a,x)}\{y-\mu_0(z,w,a,x)\} \nonumber\\
                                  &\hphantom{=}\quad+ \frac{s}{\alpha_0}q_{0}(z,x)\{\mu_0(z,w,1,x)-\mu_0(z,w,0,x)-h_{0}(w,x)\} \nonumber\\
                                  &\hphantom{=}\quad + \frac{1-s}{\alpha_0}\{h_{0}(w,x)-\theta_0\} \nonumber\\
                                  &\hphantom{=}\quad + s\{a-e_0(1\mid x)\}c(z,w,x). \label{eqn:trailing}
  \end{align}
  We now optimize over \(c\).
  Observe that the trailing term \eqref{eqn:trailing} is orthogonal to all other terms of \(\tilde{\phi}_0^{\c}(o;c,-b^*)\).
  The optimal solution is \(c^*=0\).
  From this we conclude that the efficient influence function \(\varphi^{\c}_{0}(o)=\tilde{\phi}_0^{\c}(o;c^*,-b^*)\) is as stated in Proposition~\ref{ppn:eif-mar}. 
\end{proof}

In \S\ref{sec:asymptotics}, the estimation of the outcome bridge with no missing outcome does not involve any nuisance parameter, in that the propensity score is assumed to be known, and \(\tilde{Y}_0=(2A-1)Y/e_0(A\mid X)\) can be treated as the de facto outcome in the analysis.
However, in the presence of missingness on the outcome, we resort to two-stage estimation of the outcome bridge function.
In the first stage a regression model is fitted for the outcome on the nonmissing participants, so that \(\hat{\mu}\) is an estimator for \(\mu_0\).
Additionally, we fit a binary regression model \(\hat{\pi}\) for the probability of nonmissingness \(\pi_0\).
In the second stage, the outcome bridge is estimated by a minimax optimization problem based on
\begin{multline*}
  \psi_{h',q',\pi',\mu'}(o)=q'(z,x)\bigg[\frac{2a-1}{e_0(a\mid x)}\frac{\delta}{\pi'(z,w,a,x)}\{y-\mu'(z,w,a,x)\}\\
  +\mu'(z,w,1,x)-\mu'(z,w,0,x)-h'(w,x)\bigg].
\end{multline*}
The intuition is that \(\psi_{h',q',\pi',\mu'}\) can be used to construct a doubly-robust estimating equation, in the sense that for any \(q'\), the population mean \(\E_{P^{\c}}(\psi_{h_{0},q',\pi',\mu'}\mid S=1)\) evaluated at the true outcome bridge \(h_0\) is zero if either \(\mu'=\mu_0\) or \(\pi'=\pi_0\). 
The nuisance models are plugged into \(\psi_{h',q',\pi',\mu'}\), giving the estimated outcome bridge
\[
  \hat{h}= \arg\inf_{h'\in\H'}\ \sup_{q'\in\Q'}\bigg\{\frac{1}{n_{1}}\sum_{i:S_{i}=1}\psi_{h',q',\hat{\pi}, \hat{\mu}}(O_{i})\bigg\}^{2},
\]
where \(\H'\) is the bridge hypothesis class and \(\Q'\) is the critic class \citepsuppmat{kallus2022causal}.
The former is the postulated model for the outcome bridge, and the latter is the adversarial class of functions used to construct the worst-case loss.
With the nuisance models, we compose an estimator for the target parameter \(\theta_0\) motivated by the efficient influence function \(\varphi^{\c}_{0}\) from Proposition~\ref{ppn:eif-mar}:
\[
  \hat{\theta}^{\c} = \frac{1}{n}\sum_{i=1}^{n}\bigg\{\frac{S_{i}}{\hat{\alpha}}\psi_{\hat{h},\hat{q},\hat{\pi},\hat{\mu}}(O_{i})+\frac{1-S_{i}}{\hat{\alpha}}\hat{h}(W_{i},X_{i})\bigg\}.
\]

In Theorem~\ref{thm:asymptotics-mar}, we show that the estimator \(\hat{\theta}^{\c}\) is multiply-robust under some regularity conditions and convergence of the nuisance models.
We use \(\pi_{0,a}\), \(\bar{\pi}_a\), \(\hat{\pi}_a\), \(\mu_{0,a}\), \(\bar\mu_{a}\), and \(\hat{\mu}_a\) as shorthand notations for the respective functions fixing the corresponding argument at \(a\).

\begin{assumption}[Regularity conditions]
  \label{asn:asymptotics-mar}
  \hfill
  \begin{enumerate}[label=(\roman*)]
  \item The function class
    \begin{multline*}
      \mathcal{G}_{0}^{\c}=\bigg\{g^{\c}(o)=\frac{s}{\alpha'}q'\frac{\delta(2a-1)}{\pi'e}(y-\mu')+\frac{s}{\alpha'}q'(\mu_{1}'-\mu_{0}'-h')+\frac{1-s}{\alpha'}h':\\
      \alpha'\in[0,1],q'\in L_{2}(Z,X;P^1_{0}),\pi',\mu'\in L_{2}(Z,W,A,X;P^1_{0}),h'\in L_{2}(W,X;P^1_{0})\bigg\}
    \end{multline*}
    is \(P^{\c}_0\)-Donsker.
  \item There exists a universal constant \(M>1\) such that \(\alpha_0\geq M^{-1}\), \(\hat{\alpha}\geq M^{-1}\), \(e_{0}\geq M^{-1}\), \(\pi_0\geq M^{-1}\), \(\hat{\pi}\geq M^{-1}\), \(|\mu_0|\leq M\), \(|h_{0}|\leq M\), \(|\hat{q}|\leq M\), \(P_{0}(S=1\mid W,X)\geq M^{-1}\), and \(\E_{P_{0}^\c}(Y^{2}\mid \Delta=1,Z,W,A,X,S=1)\leq M\).
    \end{enumerate}
\end{assumption}

\begin{theorem}
  \label{thm:asymptotics-mar}
  Suppose Assumption~\ref{asn:asymptotics-mar} holds and that \(\|\hat{\mu}_a-\bar{\mu}_a\|_{P_{0}^{\c,1}}=o_{P_0^{\c}}(1)\), \(\|\hat{\pi}_{a}-\bar{\pi}_{a}\|_{P_{0}^{\c,1}}=o_{P_0^{\c}}(1)\), \(\|\hat{h}-\bar{h}\|_{P_{0}^{\c,1}}=o_{P^{\c}_0}(1)\), \(\|\hat{q}-\bar{q}\|_{P_{0}^{\c,1}}=o_{P^{\c}_0}(1)\) for some nonrandom functions \(\bar{\mu}_{a}(z,w,x)\), \(\bar{\pi}_{a}(z,w,x)\), \(\bar{h}(w,x)\) and \(\bar{q}(z,x)\) in \(L_{2}(P_{0}^{\c,1})\).
  Then:
  \begin{enumerate}
  \item The estimator \(\hat{\theta}^{\c}\) is consistent for \(\theta_0\), if either (a) \(\bar{h}=h_{0}\) and \(\bar{\mu}_{a}=\mu_{0,a}\), (b) \(\bar{h}=h_{0}\) and \(\bar{\pi}_{a}=\pi_{0,a}\), (c) \(\bar{q}=q_{0}\) and \(\bar{\pi}_{a}=\pi_{0,a}\), or (d) \(\bar{q}=q_{0}\) and \(\bar{\mu}_{a}=\mu_{0,a}\).
  \item The estimator \(\hat{\theta}^{\c}\) is asymptotically linear with influence function \(\varphi^{\c}_{0}\),
    if (a) \(\bar{h}=h_{0}\), \(\bar{q}=q_{0}\), \(\bar{\pi}_{a}=\pi_{0,a}\), \(\bar{\mu}_{a}=\mu_{0,a}\), and (b) \(\|\hat{q}-q_{0}\|_{P_{0}^{\c,1}}\|\hat{h}-h_{0}\|_{P_{0}^{\c,1}}+\sum_{a\in\{0,1\}}\|\hat{\pi}_{a}-\pi_{0,a}\|_{P_{0}^{\c,1}}\|\hat{\mu}_{a}-\mu_{0,a}\|_{P_{0}^{\c,1}}=o_{P^{\c}_0}(n^{-1/2})\).
  \end{enumerate}
\end{theorem}

\begin{proof}
  Define
  \begin{align*}
    \ell^{\c}_0(o) &= \frac{s}{\alpha_0}q_{0}\frac{\delta(2a-1)}{\pi_0 e_0}(y-\mu_0)+\frac{s}{\alpha_0}q_{0}(\mu_{0,1}-\mu_{0,0}-h_{0})+\frac{1-s}{\alpha_0}h_{0},\\
    \hat{\ell}^{\c}(o) &= \frac{s}{\hat{\alpha}}\hat{q}\frac{\delta(2a-1)}{\hat{\pi}e_0}(y-\hat{\mu})+\frac{s}{\hat{\alpha}}\hat{q}(\hat{\mu}_{1}-\hat{\mu}_{0}-\hat{h})+\frac{1-s}{\hat{\alpha}}\hat{h}.
  \end{align*}
  Both \(\ell^{\c}_0\) and \(\hat{\ell}^{\c}\) belong to the \(P^{\c}\)-Donsker class \(\mathcal{G}^{\c}_{0}\), which is also \(P^{\c}\)-Glivenko-Cantelli.
  
  We first show the consistency of \(\hat{\theta}^{\c}\).
  Consider the difference
  \begin{equation}
    \label{eqn:decomp-cons}
    \hat{\theta}^{\c} - \theta_0 =(P^{\c}_{n}-P^{\c}_0)\hat{\ell}^{\c} + \bigg(P^{\c}\hat{\ell}^{\c}-\frac{\alpha_0}{\hat{\alpha}}\theta_0\bigg)-\frac{\hat{\alpha}-\alpha_0}{\hat{\alpha}}\theta_0.
  \end{equation}
  The absolute value of the first term of \eqref{eqn:decomp-cons} is bounded by \(\sup_{g^{\c}\in\mathcal{G}^{\c}_{0}}|(P^{\c}_{n}-P^{\c}_0)g^{\c}|\overset{\mathrm{P}}{\to}0\), since \(\mathcal{G}^{\c}_{0}\) is \(P^{\c}_0\)-Glivenko-Cantelli.
  The absolute value of the second term of \eqref{eqn:decomp-cons} is
  \begin{align*}
    \MoveEqLeft\bigg|P^{\c}_0\hat{\ell}^{\c}-\frac{\alpha_0}{\hat{\alpha}}\theta_0\bigg|\\
    &= \bigg|P^{\c}_0\bigg[\frac{S}{\hat{\alpha}}\hat{q}\frac{2A-1}{e_0}\frac{\Delta}{\hat{\pi}}(Y-\hat{\mu})+\frac{S}{\hat{\alpha}}\hat{q}(\hat{\mu}_{1}-\hat{\mu}_{0}-\hat{h})+\frac{1-S}{\hat{\alpha}}\hat{h}\bigg]-\frac{\alpha_0}{\hat{\alpha}}\theta_0\bigg| \\
    &= \bigg|\frac{1}{\hat{\alpha}}P^{\c}_0\bigg[S\hat{q}\bigg\{\frac{\pi_{0,1}}{\hat{\pi}_{1}}(\mu_{0,1}-\hat{\mu}_{1})-\frac{\pi_{0,0}}{\hat{\pi}_{0}}(\mu_{0,0}-\hat{\mu}_{0})\bigg\}+S\hat{q}(\hat{\mu}_{1}-\hat{\mu}_{0}-\hat{h})+Sq_{0}\hat{h}\bigg]-\frac{\alpha_0}{\hat{\alpha}}\theta_0\bigg| \\
    &= \bigg|\frac{1}{\hat{\alpha}}P^{\c}_0\bigg\{S\sum_{a\in\{0,1\}}(-1)^{1-a}\hat{q}\frac{\pi_{0,a}-\hat{\pi}_{a}}{\hat{\pi}_{a}}(\mu_{0,a}-\hat{\mu}_{a}) + S\hat{q}(\mu_{0,1}-\hat{\mu}_{1}-\mu_{0,0}+\hat{\mu}_{0})\\
    &\hphantom{=\bigg|\frac{1}{\hat{\alpha}}P^{\c}_0\bigg\{}\quad+S\hat{q}(\hat{\mu}_{1}-\hat{\mu}_{0}-\hat{h})+Sq_{0}\hat{h}\bigg\}-\frac{\alpha_0}{\hat{\alpha}}\theta_0\bigg| \\
    &= \bigg|\frac{1}{\hat{\alpha}}P^{\c}_0\bigg\{S\sum_{a\in\{0,1\}}(-1)^{1-a}\hat{q}\frac{\pi_{0,a}-\hat{\pi}_{a}}{\hat{\pi}_{a}}(\mu_{0,a}-\hat{\mu}_{a})-S(\hat{q}-q_{0})(\hat{h}-h_{0})+Sh_{0}q_{0}\bigg\}-\frac{\alpha_0}{\hat{\alpha}}\theta_0\bigg| \\                                                                                           
    &\leq M^{3}\sum_{a\in\{0,1\}}\|\hat{\pi}_{a}-\pi_{0,a}\|_{P_{0}^{\c,1}}\|\hat{\mu}_{a}-\mu_{0,a}\|_{P_{0}^{\c,1}}+ M\|\hat{h}-h_{0}\|_{P_{0}^{\c,1}}\|\hat{q}-q_{0}\|_{P_{0}^{\c,1}}\overset{\mathrm{P}}{\to}0.
  \end{align*}
  The absolute value of the third term of \eqref{eqn:decomp-cons} is trivially \(o_{P^{\c}_0}(1)\) due to the consistency of \(\hat{\alpha}\) and Slutsky's theorem.
  Collecting these three terms, the triangular inequality shows \(|\hat{\theta}^{\c}-\theta_0|=o_{P^{\c}}(1)\).
  This shows the first part of the theorem.
  
  We now show the asymptotic linearity of \(\hat{\theta}^{\c}\).
  Working under the additional assumption \(\sum_{a\in\{0,1\}}\|\hat{\pi}_{a}-\pi_{0,a}\|_{P_{0}^{\c,1}}\|\hat{\mu}_{a}-\mu_{0,a}\|_{P_{0}^{\c,1}}+\|\hat{h}-h_{0}\|_{P_{0}^{\c,1}}\|\hat{q}-q_{0}\|_{P_{0}^{\c,1}}=o_{P^{\c}_0}(n^{-1/2})\), we further express the difference as
  \begin{align*}
     \hat{\theta}^{\c} - \theta_0 &= (P_{n}^{\c}-P^{\c}_0)(\hat{\ell}^{\c}-\ell^{\c}_0) + P^{\c}_{n}\ell^{\c}_0 - \theta_0 + \bigg(P^{\c}_0\hat{\ell}^{\c}-\frac{\alpha_0}{\hat{\alpha}}\theta_0\bigg)-\frac{\hat{\alpha}-\alpha_0}{\hat{\alpha}}\theta_0 \\
     &= P_{n}^{\c}\varphi^{\c}_{0}+ (P_{n}^{\c}-P^{\c}_0)(\hat{\ell}^{\c}-\ell^{\c}_0)+\frac{(\hat{\alpha}-\alpha_0)^{2}}{\alpha_0\hat{\alpha}}\theta_0+o_{P^{\c}_0}(n^{-1/2}).
  \end{align*}
  The second term is an empirical process term of order \(o_{P^{\c}_0}(n^{-1/2})\) if \(\|\hat{\ell}^{\c}-\ell^{\c}_0\|_{P^{\c}_0}=o_{P^{\c}_0}(1)\), since \(\mathcal{G}^{\c}_{0}\) is \(P^{\c}_0\)-Donsker.
  By an application of the central limit theorem to \(\hat{\alpha}\) and Slutsky's theorem, the third term is of the order \(O_{P^{\c}_0}(n^{-1})=o_{P^{\c}_0}(n^{-1/2})\).
  To conclude the proof of the second part of the theorem, we show that \(\|\hat{\ell}^{\c}-\ell^{\c}_0\|_{P^{\c}_0}\) indeed converges in probability to zero.
  The \(L_{2}(P^{\c}_0)\)-norm of the plugin function \(\hat{\ell}^{\c}\) is
  \begin{align*}
    \|\hat{\ell}^{\c}\|_{P^{\c}_0} &\leq M\bigg\{\bigg\|S\hat{q}\frac{(2A-1)}{e_0(A\mid X)}\frac{\Delta}{\hat{\pi}}Y\bigg\|_{P^{\c}_0}+\bigg\|S\hat{q}\frac{(2A-1)}{e_0(A\mid X)}\frac{\Delta}{\hat{\pi}}\hat{\mu}\bigg\|_{P^{\c}_0}\\
                                   &\hphantom{\leq M\bigg\{}\quad+\|S\hat{q}(\hat{\mu}_{1}-\hat{\mu}_{0}-\hat{h})\|_{P^\c_0}+\|(1-S)\hat{h}\|_{P^\c_0}\bigg\} \\
                                       &\leq M^{4}\{\E_{P^{\c}}(Y^{2}\mid \Delta=1,S=1)\}^{1/2}+M^{7/2}\sum_{a\in\{0,1\}}\|\hat{\mu}_{a}\|_{P_{0}^{\c,1}}\\
    &\hphantom{\leq}\quad +\bigg\{M^{2}\sum_{a\in\{0,1\}}\|\hat{\mu}_{a}\|_{P_{0}^{\c,1}}+M^{2}\|\hat{h}\|_{P_{0}^{\c,1}}\bigg\}+M^{3/2}\|\hat{h}\|_{P_{0}^{\c,1}} \\
    &=O_{P^{\c}_0}(1).
  \end{align*}
  This is because the norms of the nuisance estimators \(\|\hat{q}\|_{P_{0}^{\c,1}}\leq M=O_{P_{0}^{\c}}(1)\), \(\|\hat{h}\|_{P_{0}^{\c,1}}\leq \|\hat{h}-\bar{h}\|_{P_{0}^{\c,1}}+\|\bar{h}\|_{P_{0}^{\c,1}}=O_{P_{0}^{\c}}(1)\) and \(\|\hat{\mu}_{a}\|_{P_{0}^{\c,1}}\leq \|\hat{\mu}_a-\bar{\mu}_a\|_{P_{0}^{\c,1}}+\|\bar{\mu}_a\|_{P_{0}^{\c,1}}=O_{P_{0}^{\c}}(1)\) are bounded by probability.
  The \(L_{2}(P^{\c}_0)\)-distance between the plugin and the true function is
  \begin{align*}
    \MoveEqLeft\|\hat{\ell}^{\c}-\ell^{\c}_0\|_{P^{\c}_0} \\
    &\leq \frac{1}{\alpha_0}\bigg\|S(\hat{q}-q_{0})\frac{2A-1}{e_0}\frac{\Delta}{\pi_0}Y\bigg\|_{P^{\c}_0} + \frac{1}{\alpha_0}\bigg\|S\hat{q}\frac{2A-1}{e_0}\frac{\Delta(\pi_0-\hat{\pi})}{\hat{\pi}\pi_0}Y\bigg\|_{P^{\c}_0}\\
                                                        &\hphantom{\leq}\quad + \frac{1}{\alpha_0}\bigg\|S(\hat{q}-q_{0})\frac{2A-1}{e}\frac{\Delta}{\pi_0}\mu_0\bigg\|_{P^{\c}_0} + \frac{1}{\alpha_0}\bigg\|S\hat{q}\frac{2A-1}{e_0}\frac{\Delta(\pi_0-\hat{\pi})}{\hat{\pi}\pi_0}\mu\bigg\|_{P^{\c}_0}\\
                                             &\hphantom{\leq}\quad + \frac{1}{\alpha_0}\bigg\|S\hat{q}\frac{2A-1}{e_0}\frac{\Delta}{\hat{\pi}}(\hat{\mu}-\mu_0)\bigg\|_{P^{\c}_0}+ \frac{1}{\alpha_0}\|S\{\hat{q}(\hat{\mu}_{1}-\hat{\mu}_{0}-\hat{h})-q_0(\mu_{0,1}-\mu_{0,0}-h_{0})\}\|_{P^{\c}_0}\\
                                                        &\hphantom{\leq}\quad +\frac{1}{\alpha_0}\|(1-S)(\hat{h}-h_{0})\|_{P^{\c}_0}+\frac{|\hat{\alpha}-\alpha_0|}{\alpha_0}\|\hat{\ell}^{\c}\|_{P^{\c}_0},\\
    \intertext{and by similar arguments above, we bound the distance by}
                                                        &\leq M^{3}[P^{\c}_0\{S(\hat{q}-q_{0})^{2}\E_{P^{\c}_0}(Y^{2}\mid \Delta=1,Z,X,S=1)\}]^{1/2} \\
                                                        &\hphantom{=}\quad + M^{4}\sum_{a\in\{0,1\}}[P^{\c}_0\{S(\hat{\pi}_{a}-\pi_{0,a})^{2}\E_{P^{\c}_0}(Y^{2}\mid \Delta=1,Z,W,A=a,X,S=1)\}]^{1/2}\\
                                                        &\hphantom{=}\quad + M^{3}\|\hat{q}-q_{0}\|_{P_{0}^{\c,q}}+M^{5}\sum_{a\in\{0,1\}}\|\hat{\pi}_{a}-\pi_{0,a}\|_{P_{0}^{\c,1}} + M^{7/2}\sum_{a\in\{0,1\}}\|\hat{\mu}_{a}-\mu_{0,a}\|_{P_{0}^{\c,1}} \\
    &\hphantom{\leq}\quad +\{3M^{2}\|\hat{q}-q_{0}\|_{P_0^{\c,1}}+M^{2}\|\hat{\mu}_{1}-\mu_{0,1}\|_{P_{0}^{\c,1}}+M^{2}\|\hat{\mu}_{0}-\mu_{0,0}\|_{P_{0}^{\c,1}}+M^{2}\|\hat{h}-h_{0}\|_{P_{0}^{\c,1}}\}\\
                                                        &\hphantom{\leq}\quad +M^{3/2}\|\hat{h}-h_{0}\|_{P_{0}^{\c,1}}+ M|\hat{\alpha}-\alpha_0|O_{P^{\c}_0}(1)\\
    &=o_{P^{\c}_0}(1).
  \end{align*}
  We conclude that \(\hat{\theta}^{\c}-\theta_0=P_{n}^{\c}\varphi^{\c}_{0}+o_{P^{\c}_0}(n^{-1/2})\).
\end{proof}

\subsection{Simulation}
In the simulation study for the indirect comparison estimator in the presence of missing outcomes, we generated the full data \((S\Delta,U,S,X,SA,Y,W,SZ)\) including a missing indicator \(\Delta\).
We sampled from the distribution of \((U,S,X,SA,Y,W,SZ)\) specified in \S\ref{sec:simulation} and drew \(\Delta\) from the distribution
\[
  \Delta\mid (Z,W,A,X,S=1) \sim \mathrm{Bernoulli}\{\mathrm{expit}(0.1Z^{\T}1+0.1W^{\T}1+0.7A+0.3X^{\T}1)\}.
\]
We only investigated the multiply robust estimator \(\hat{\theta}^{\c}\).
On the source RCT sample, we fitted the adherence probability model \(\hat{\pi}(z,w,a,x)\) using a logistic regression linear in all covariates and the mean outcome model \(\hat{\mu}(z,w,a,x)\) on the subsample where \(\Delta=1\) using an ordinary linear regression with interaction between \(A\) and \((Z,W,X)^{\T}\).
The nuisance estimator of the outcome difference bridge was subsequently obtained as
\[
  \hat{\eta}=\arg\min_{\eta'}\bigg\|\frac{1}{n_1}\sum_{i:S_i=1}\psi_{h_{\eta'},b,\hat{\pi},\hat{\mu}}(O_{i})\bigg\|^{2},
  \]
  while the estimation of the participation odds bridge remained the same as in the setup without nonadherence.
To demonstrate the robustness against model misspecifications, we considered the configurations where none of \(h\), \(q\), \(\pi\), and \(\mu\) was misspecified (experiment 14), where \(q\) and \(\pi\) were misspecified (experiment 15), where \(q\) and \(\mu\) were misspecified (experiment 16), where \(h\) and \(\pi\) were misspecified (experiment 17), where \(h\) and \(\mu\) were misspecified (experiment 18) and where all of \(h\), \(q\), \(\pi\), and \(\mu\) were misspecified (experiment 19).
The misspecified models were fitted by replacing \(W\) and \(Z\) with \(|W|^{1/2}\) and \(|Z|^{1/2}\) wherever appropriate.
However, the true model \(\mu_0\) for \(\mu\) does not have an easy closed-form expression.
Therefore, all the posited models for \(\mu\) could have been misspecified, whether intentionally or not.
The results are displayed in Table~\ref{tab:sim-mar}.
The estimator \(\hat{\theta}^{\c}\) retained small empirical biases under model misspecifications as expected, except when all nuisance models were misspecified.
The influence-function-based standard error in experiments \(15\)--\(16\) where the \(q\) model was misspecified led to anticonservative confidence intervals.

\begin{table}
  \caption{Simulation results of experiments 14--19.}
  \label{tab:sim-mar}
  \footnotesize\centering
  {\begin{tabular}[t]{llS[]S[]rrr}
    \toprule
    {\(n\)} & {Experiment} & {Mean} & {Bias} & {RMSE} & {SE} & {Coverage}\\
    \midrule
    \(1000\) & 14 & -2.64 & 1.57 & 1.19 & 1.15 & 93.9\\
            & 15 & -2.64 & 0.33 & 1.24 & 1.12 & 91.8\\
            & 16 & -2.64 & 3.08 & 3.08 & 2.68 & 90.7\\
            & 17 & -2.64 & 5.97 & 1.24 & 1.27 & 95.7\\
            & 18 & -2.63 & 11.51 & 3.30 & 3.16 & 94.0\\
            & 19 & -2.40 & 248.88 & 3.73 & 2.72 & 82.3\\
    \midrule
    \(2000\) & 14 & -2.65 & -0.36 & 0.84 & 0.81 & 94.1\\
            & 15 & -2.65 & -0.24 & 0.84 & 0.77 & 93.1\\
            & 16 & -2.65 & -5.38 & 2.08 & 1.85 & 91.7\\
            & 17 & -2.64 & 2.67 & 0.86 & 0.88 & 95.8\\
            & 18 & -2.65 & -3.09 & 2.23 & 2.22 & 95.1\\
            & 19 & -2.39 & 250.89 & 3.15 & 1.85 & 72.8\\
    \bottomrule
  \end{tabular}}

\medskip
  {Bias: Monte-Carlo bias, \(10^{-3}\); RMSE: root mean squared error, \(10^{-1}\); SE: average of standard error estimates, \(10^{-1}\); Coverage: \(95\%\) confidence interval coverage, \(\%\).}
\end{table}

\bibliographysuppmat{bibliography}


\end{document}